\documentclass[aps,prd,twocolumn,showpacs,amsmath,amssymb,nofootinbib]{revtex4-1}
\usepackage{graphicx}
\usepackage{color}
\usepackage{times}
\usepackage{inputenc}
\usepackage{bm}
\usepackage{ulem}
\usepackage{multirow}
\usepackage{float}
\usepackage{url}
\usepackage{natbib}
\usepackage{mathrsfs}
\usepackage{physics}
\usepackage{comment}
\usepackage[table,xcdraw]{xcolor}
\usepackage{booktabs}
\usepackage{comment}
\usepackage[caption=false]{subfig}
\usepackage[colorlinks=true,citecolor=blue,urlcolor=blue,linkcolor=blue]{hyperref}

\newcommand{\aap}{A\&A} 

\usepackage{tikz,xcolor,hyperref}
\definecolor{lime}{HTML}{A6CE39}
\DeclareRobustCommand{\orcidicon}{%
	\begin{tikzpicture}
	\draw[lime, fill=lime] (0,0) 
	circle [radius=0.16] 
	node[white] {{\fontfamily{qag}\selectfont \tiny ID}};
	\draw[white, fill=white] (-0.0625,0.095) 
	circle [radius=0.007];
	\end{tikzpicture}
	\hspace{-2mm}
}
\foreach \x in {A, ..., Z}{%
	\expandafter\xdef\csname orcid\x\endcsname{\noexpand\href{https://orcid.org/\csname orcidauthor\x\endcsname}{\noexpand\orcidicon}}
}

\begin{document}
\title{Unstable Anisotropic Neutron Stars: Probing the Limits of Gravitational Collapse}
\author{Sailesh Ranjan Mohanty$^{1}$\orcidA{}} 
\author{Sayantan Ghosh$^{1}$\orcidB{}}
\author{Bharat Kumar$^{1}$\orcidC{}}
\email{kumarbh@nitrkl.ac.in}
\affiliation{\it $^{1}$Department of Physics and Astronomy, National Institute of Technology, Rourkela 769008, India}
\date{\today}
\begin{abstract}
{Neutron stars (NSs) are incredibly versatile for studying various important aspects of high-energy and compact-object physics. These celestial objects contain extreme matter at incredibly high densities in their interiors, leading to the risk of instabilities that may cause them to collapse into a black hole (BH). This paper focuses on exploring the stability and gravitational collapse of NSs. For a more realistic approach we have considered the pressure to be locally anisotropic. We utilize the BL-Model to describe the anisotropy inside the NS. The presence of quarks in the core of an NS can heavily affect its stability. Hence, along with pure hadronic EOSs, we have also considered Hadron-Quark phase transition (HQPT) EOSs for this paper's analysis. We subject the anisotropic NSs to radial perturbations to study their stability against radial oscillations. NSs exhibiting imaginary eigen-frequencies are identified as unstable, and their inevitable destiny is gravitational collapse, resulting in the formation of a BH. We consider the interior of these unstable anisotropic NSs to be a non-ideal fluid in a non-adiabatic background in order to study its dynamical evolution during the collapse. We examine the temporal evolution of key properties of NSs, such as mass, density, heat flux, and anisotropy during the process of gravitational collapse. We present an innovative and viable approach to detect such high-energy gravitational collapse events, providing valuable insights into the properties of the static NS before its collapse.}
\end{abstract}
\maketitle
\section{Introduction}
\label{intro}
Understanding the formation and evolution of compact stars, such as neutron stars (NSs), is a challenging task, requiring expertise in various areas of physics. Despite its complex internal structure and strong gravitational field with extreme matter, a complete theoretical understanding of this object is yet to be achieved \cite{Lattimer}. An isotropic perfect fluid is the most common matter-energy distribution used to model the interior structure of compact stars. However, this has strong arguments against it, as nuclear matter at very high pressures and densities may cause deviations in the tangential and radial components of the pressure, resulting in an anisotropic fluid. There are numerous factors that can generate anisotropy in NSs, including the presence of a high magnetic field \cite{S.S.Yaza,C.Y.Cardall_2001,K.Ioka_2004,R.Ciolfi_L.Rezzolla,R.Ciolfi_V.Ferrari,Frieben,A.G.Pili,N.Bucciantini}, pion condensation \cite{R.F.Sawyer}, phase transitions \cite{B.Carter}, relativistic nuclear interaction \cite{V.Canuto,M.Ruderman}, core crystallization \cite{S.Nelmes}, and superfluid cores \cite{W.A.Kippenhahn_Rudolf,Glendenning:1997wn,H.Heiselberg}, among others \cite{Reasons_of_anisotropy}. Moreover, an anisotropic perfect fluid can be created by combining the energy-momentum tensors of two isotropic perfect fluids. \cite{Letelier,2013rehy.book.....R,Reasons_of_anisotropy}.

Anisotropic models have been proposed in the literature that maintain a perfect fluid arrangement, including the Bowers-Liang (BL) \cite{BL-Model}, Cosenza et al. \cite{Cosenza}, and Horvat et al. \cite{Horvat_2011} models. In this study, we will focus on the BL model to describe the anisotropy inside NSs as it is quadratically driven by gravity and relies nonlinearly on radial pressure. Anisotropy has a significant impact on many macroscopic characteristics of compact stars, such as the mass-radius relation, compactness, surface redshift, moment of inertia, and tidal deformability \cite{Pretel2020,Hillebrandt,singh2022study,Silva_2015,Doneva,Bayin,Roupas2021,Deb_2021,Estevez-Delgado2018,Pattersons2021,Rizaldy,Rahmansyah2020,Rahmansyah2021,Herrera2008,Herrera2013,Bhaskar_Biswas,S.Das2021,Roupas2020,Sulaksono2015,Setiawan2019,Silva_2015}. The presence of anisotropy is mainly determined by the equation of state (EOS), which imposes stability restrictions and determines the interior structure of these stars. It is worth noting that LIGO-Virgo constraints on the EOS for nuclear matter should be taken into account as they provide important information about the properties of dense matter \cite{GW170817}. In this study, we investigate the effect of a plausible EOS that fits the constraints from the GW170817 merger on the physical properties of both stable and unstable anisotropic stars.

The Tolman-Oppenheimer-Volkoff (TOV) equations describe the structure of stars in hydrostatic equilibrium, but it's important to note that this state of equilibrium can be either stable or unstable. The limit that separates stable and unstable stars is where a given stellar configuration may collapse under the force of gravity. One of the most common methods to determine the stability of a NS is by requiring that its mass must increase with increasing central density ($\frac{\partial M}{\partial \rho_c } > 0 $). This is part of the Bardeen-Thorne-Meltzer \cite{Bardeen} stability criterion, also known as the $M(\rho_c)$ method, which is also used by authors in \cite{Horvat_2011}. Although violations of this criterion have been known to exist numerically for decades \cite{Gourgoulhon_1995}, it is still often used as it only requires solving the TOV equation for a given EOS. A more fundamental criterion for stability is that the radial modes of oscillation of a NS must have real frequencies \cite{Glendenning:1997wn,Haensel:2007yy,Kokkotas}. The NS is said to be stable if the fundamental mode ($f$-mode) has a positive frequency squared ($\omega^2>0$). The cause of these oscillations can be fluid pulsations, which may result from cracks in the crust, also known as ``starquakes" \cite{Franco_2000}, or tidal effects from binary inspirals \cite{Hinderer2016,Chirenti_2017}. One can also obtain critical central densities ($\rho_c =\Bar{\rho}_c$) corresponding to the maximally stable neutron star where the minimum stability condition (also known as the stability limit) is satisfied. NSs with central densities beyond the critical density are considered to be unstable and are expected to collapse. Recent literature has shown that the $M(\rho_c)$ method cannot be used in conjunction with the radial stability criterion to predict the onset of instability in anisotropic NS systems \cite{Horvat_2011,Pretel2020,Hillebrandt,singh2022study}. This is because there is no correspondence between the stability limit obtained from the squared frequencies of the fundamental mode $\left( \omega^2 (\bar{\rho}_{c}) = 0\right)$ and the star configurations that produce the maximum mass $\left( \left. \frac{\partial M}{\partial \rho_c } \right|_{\bar{\rho}_{c}} = 0 \right)$. Importantly, these studies have underscored that the stability limits can diverge between isotropic and anisotropic NSs. Additionally, there's a possibility for an initially isotropic NS to abandon its isotropic nature during a dissipative gravitational collapse, generating anisotropy in the pressure \cite{Herrera_2020}. Hence, the stability analysis of NSs requires careful consideration of their anisotropy.

Apart from anisotropic NSs, extensive research on the stability analysis with respect to radial perturbations has also been conducted for rotating stars \cite{Takami,Weih,staykov2023}, strange stars \cite{10.1093/mnras/250.4.679,Arbanil_2016,C.V_Flores,Panotopoulos_2017,Panotopoulos_2018}, hybrid stars \cite{Pereira_2019}, dark stars \cite{PhysRevD.107.103039,routaray2023probing,Adrian_2023,Gresham_2019}, and many more \cite{Kokkotas,Chandrasekhar,Chanmugam,Sagun_2020,Clemente,Dev2003,Isayev}, leading to a substantial body of literature on this subject. Hybrid stars are one of the most promising candidates for realistic NSs. These types of NSs follow a hadron-quark phase transition (HQPT) EOS. In the core of such stars, where the density is beyond the nuclear density, a phase transition called quark deconfinement may occur. This involves the conversion of neutrons into quark matter, where quarks are no longer confined within neutrons but can move more freely, and hence significantly affects the stability of hybrid NSs. The study of stellar oscillations of compact stars with phase transitions was pioneered by \cite{Kogan_1984,Haensel_1989} for nonrelativistic stellar models. These studies highlighted how fast the change from one phase to another happens. Whether a fluid keeps its phase during oscillation across the equilibrium boundary depends on how quickly this change occurs compared to the oscillation period. Authors in ref. \cite{Pereira_2018,Rau_2023,Goncalves_2022,PhysRevD.108.103035, ghosh2024exploring,VasquezFlores_2012,Gupta_2002} have demonstrated that in the context of hybrid NSs featuring a phase transition EOS, the conventional criteria associated with the $M(\rho_c)$ method may not hold true. The rate of a phase transition may additionally have physically relevant implications for the stability of compact stars. They found that stars exhibiting $\partial M/\partial \rho_c < 0$ can remain stable against linearized radial perturbations, provided that the hadron-quark conversion occurs at a rate slower than the timescale of the perturbation.  Such stars were coined as slow-stable hybrid stars. Hence, the most natural approach for stability analysis is by means of radial perturbations, which is the route we take in this work.

As stated earlier, the ultimate fate of an unstable compact star is to collapse into a black hole under the force of its own gravity. Designing a physically valid gravitational collapse scenario is quite challenging. One of the earliest studies of gravitational collapse was conducted by Oppenheimer and Snyder on spherically symmetric systems of dust clouds \cite{Oppenheimer1939}. This idealized method was later improved by introducing a pressure gradient and switching from the external Schwarzschild metric to the Vaidya metric \cite{Misner1964,Vaidya1951}. Physically valid gravitational collapse scenarios for isotropic fluids were developed by incorporating dissipative fluxes {\cite{Misner1965,N.O.Santos,A.K.G._de_Oliveira,L.Herrera1989,W.B.Bonnor,L.Herrera_2006,B.V.Ivanov2012,J.Martinez_1994,A.K.G.deOliveira}}. Moreover, dynamical models have been developed to describe gravitational collapse in anisotropic fluids with dissipative processes \cite{J.Martinez,2015GReGr..47...35R,L.S.M.Veneroni,J.M.Z.Pretel,M.Govender}, electromagnetic fields \cite{A.DiPrisco,G.Pinheiro,Ivanov_2019,M.Z.Bhatti}, and cosmological constants \cite{A.B.Mahomed}. Additionally, the stability of initial static Schwarzschild configurations in relation to radial pulsations has been studied \cite{Pretel,Pretel2020}.

In this paper, we aim to extend the study of the stability of compact stars with two-phase transitions in an anisotropic environment and also examine the dynamics behind the gravitational collapse of the resulting unstable NSs. A similar type of research has been conducted in ref. \cite{Pretel2020}, but investigation of stability of anisotropic hybrid NSs has not been carried out in any previous literature. Along with soft and stiff variants of sharp HQPT EOS \cite{demircik2021dense}, we use two hadronic EOSs (BSk21 \cite{BSK21} and SLY4 \cite{SLY4}) which are consistent with recent observational data. First, we construct a family of anisotropic NSs using the above-mentioned EOSs and explore their properties. For stability analysis, we investigate their oscillation spectra by applying radial perturbations. Concerning perturbations in hybrid stars, one cannot integrate the system of equations as done in one-phase stars. Instead, careful consideration must be given to additional boundary conditions at the phase-transition interfaces. These interfaces essentially encapsulate the main physics occurring in their proximity. Around such interfaces, in the presence of perturbations, two primary types of physical behavior emerge: either volume elements undergo conversion from one phase to another, or they maintain their nature and experience only stretching or compression. The former is associated with rapid phase transitions, while the latter pertains to slow phase transitions \cite{Haensel_1989}. Following stability analysis, we filter out all the unstable stars and study their evolution over time, which will result in the formation of an event horizon due to non-adiabatic gravitational collapse. We assume that the unstable star comprises a non-ideal fluid exhibiting bulk viscosity, radial heat flow, and outward radiation flux throughout the process of dissipative gravitational collapse.  Additionally, we also calculate the large frequency difference (difference between consecutive nodes) and study the effects of anisotropy on it, which helps to understand the internal structure of the stars and is commonly used in asteroseismology. Finally, we discuss ideas for detecting such high-energy gravitational collapses of anisotropic stars. Throughout this paper, we adopt mostly positive signatures (-, +, +, +) and utilize geometrical units ({$G=c=1$}) 
\section{Stellar Structure Equations}
\begin{figure*}
    \centering
    \includegraphics[width=0.49\linewidth]{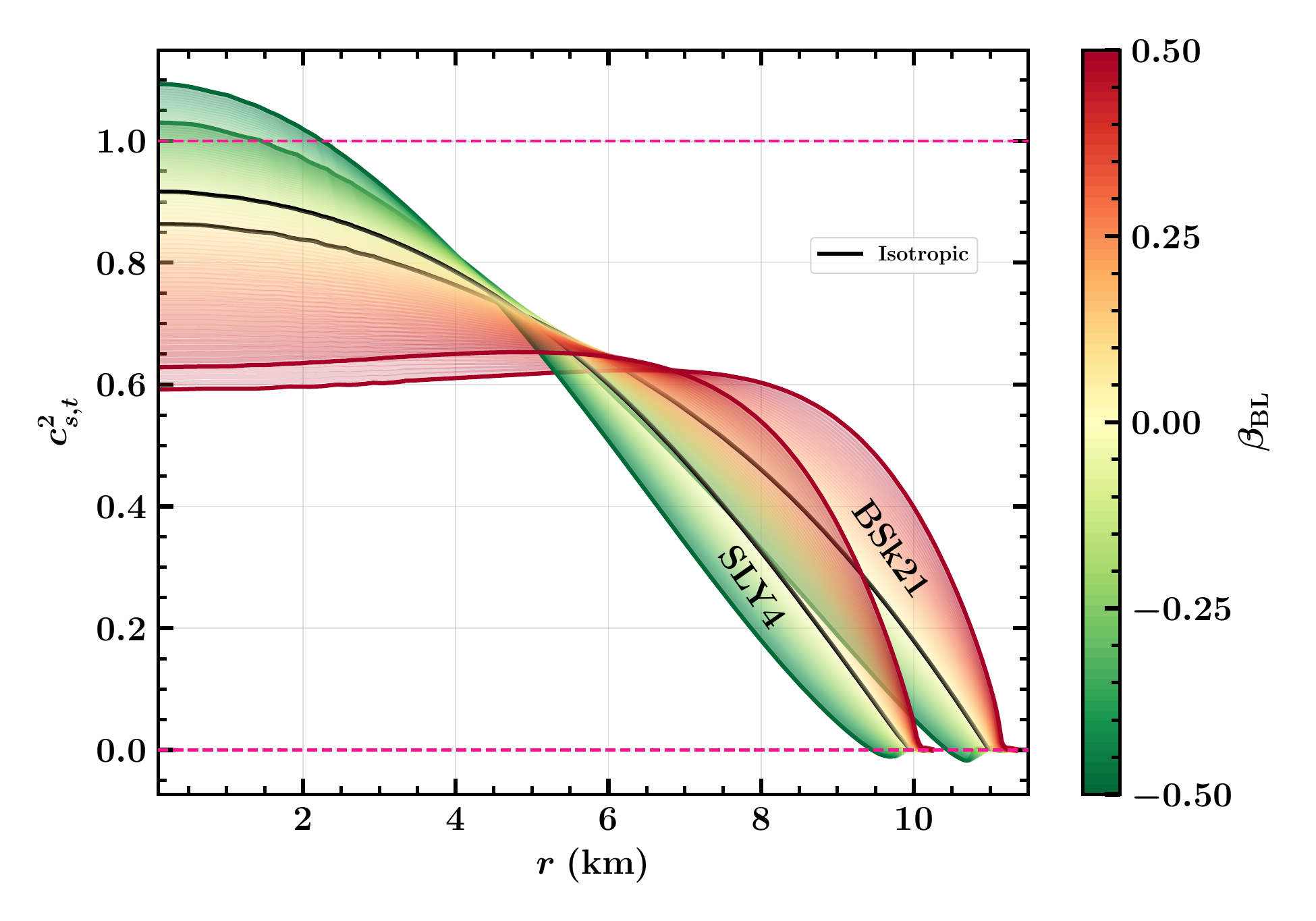}
    \includegraphics[width=0.49\linewidth]{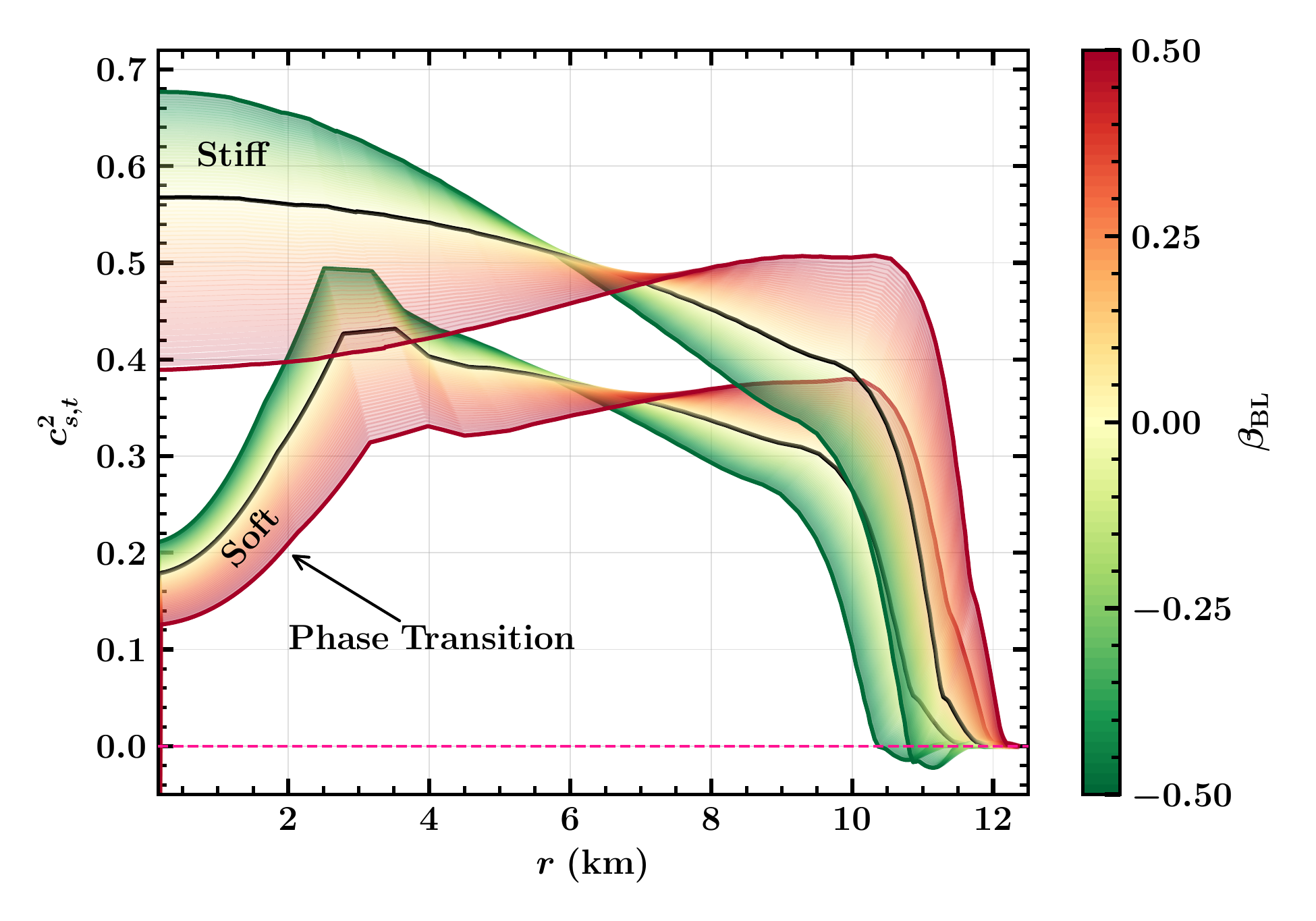}
    \caption{The radial profile of sound speed ($c_{s,t}^{2}$) for different $\beta_\mathrm{BL}$ of the maximum mass NS corresponding to respective EOSs as depicted in the picture.}
    \label{fig:Sound_speed}
\end{figure*}
The line element that describes the space-time inside a spherically symmetric  star is \cite{Wald:1984rg,Schwarzschild:1916uq}
\begin{equation}
\label{eq:metric}
ds^2=-e^{2 \psi}\left(d t\right)^2+e^{2 \lambda} d r^2+r^2 (d \theta^2+\sin ^2 \theta d \phi^2)
\end{equation}
where, $\psi$ and $\lambda$ are metric functions that depend on $r$. Introducing anisotropy in the matter-energy distribution of the system, we obtain the following stress-energy tensor \cite{Estevez-Delgado2018}
\begin{equation}
\label{eq:energy_mom}
    T_{\mu v}=\left(\mathcal{E}+P_t\right) u_\mu u_v+P_t g_{\mu v}-\sigma k_\mu k_v,
\end{equation}
where $u^\mu$ is the four-velocity of the fluid, and $k^\mu$ is a unit space-like four-vector. $\mathcal{E}$ is the energy density, and $\sigma$ is a anisotropy pressure ($\sigma = P_t - P_r$). The four-vectors $u^\mu$ and $k^\mu$ must satisfy the following
\begin{equation}
\label{eq:four_vector_conditions}
    u_\mu u^\mu=-1, \quad k_\mu k^\mu=1, \quad u_\mu k^\mu=0 .
\end{equation}
The Einstein field equations relate the space-time geometry and the matter-energy distribution \cite{Einstein:1915ca}
\begin{equation}
\label{eq:einstein}
    G_{\mu \nu} = R_{\mu \nu}-\frac{1}{2} R g_{\mu \nu}=8\pi T_{\mu \nu},
\end{equation}
with $G_{\mu \nu}$ being the Einstein tensor, $R_{\mu \nu}$ the Ricci tensor, $R$ denoting the scalar curvature, and $T_{\mu \nu}$ being the stress-energy tensor. Solving Eqs. (\ref{eq:metric}-\ref{eq:einstein}) with the energy and momentum conservation, one can obtain the modified TOV equations which describe the stellar structure of an anisotropic star given as \cite{mod_TOV,TOV}
\begin{align}
\label{eq:TOV}
    \frac{d m}{d r} &=4 \pi r^{2} \mathcal{E}, \nonumber \\ 
    \frac{d P_{r}}{d r} &=-\left({P_{r}+ \mathcal{E}}\right)\left(\frac{m}{r^{2}}+{4 \pi } r P_{r}\right) e^{2 \lambda}+\frac{2}{r} \sigma, \nonumber \\
    \frac{d \psi}{d r} &=-\frac{1}{P_{r}+ \mathcal{E}} \frac{d P_{r}}{d r}+\frac{2 \sigma}{r\left(P_{r}+ \mathcal{E}\right)},
\end{align}
where $m(r)$ is the enclosed mass corresponding to radius $r$ and $\lambda(r)$ is the metric function defined as
\begin{equation*}
    e^{-2 \lambda}=1-\frac{2 m}{r}.
\end{equation*}
To solve the above-coupled ODEs numerically, we need to set up some boundary conditions. We usually find the surface of a star at the point $(r=R)$ where the radial pressure is equal to zero. So, $P_{r}=0$ is true as long as $r \rightarrow R$. The equilibrium system is spherically symmetric, so the Schwarzschild metric must be used to describe the space-time outside the system. This ensures the continuity of the metric on the surface of the anisotropic NS, which ultimately imposes a boundary condition on $\psi$. 
\begin{equation*}
\psi(r=R)=\frac{1}{2} \ln \left[1-\frac{2  M}{ R}\right].
\end{equation*}
With a given EOS for the radial pressure ($P_r$) and an anisotropic model for $\sigma$, Eq. (\ref{eq:TOV}) can be numerically solved for a given central energy density $\mathcal{E}(r=0)=\mathcal{E}_{c}$ with $m(r=0)=0$. 
\section{Anisotropy Model}
\begin{figure*}
    \centering
    \includegraphics[width=0.49\linewidth]{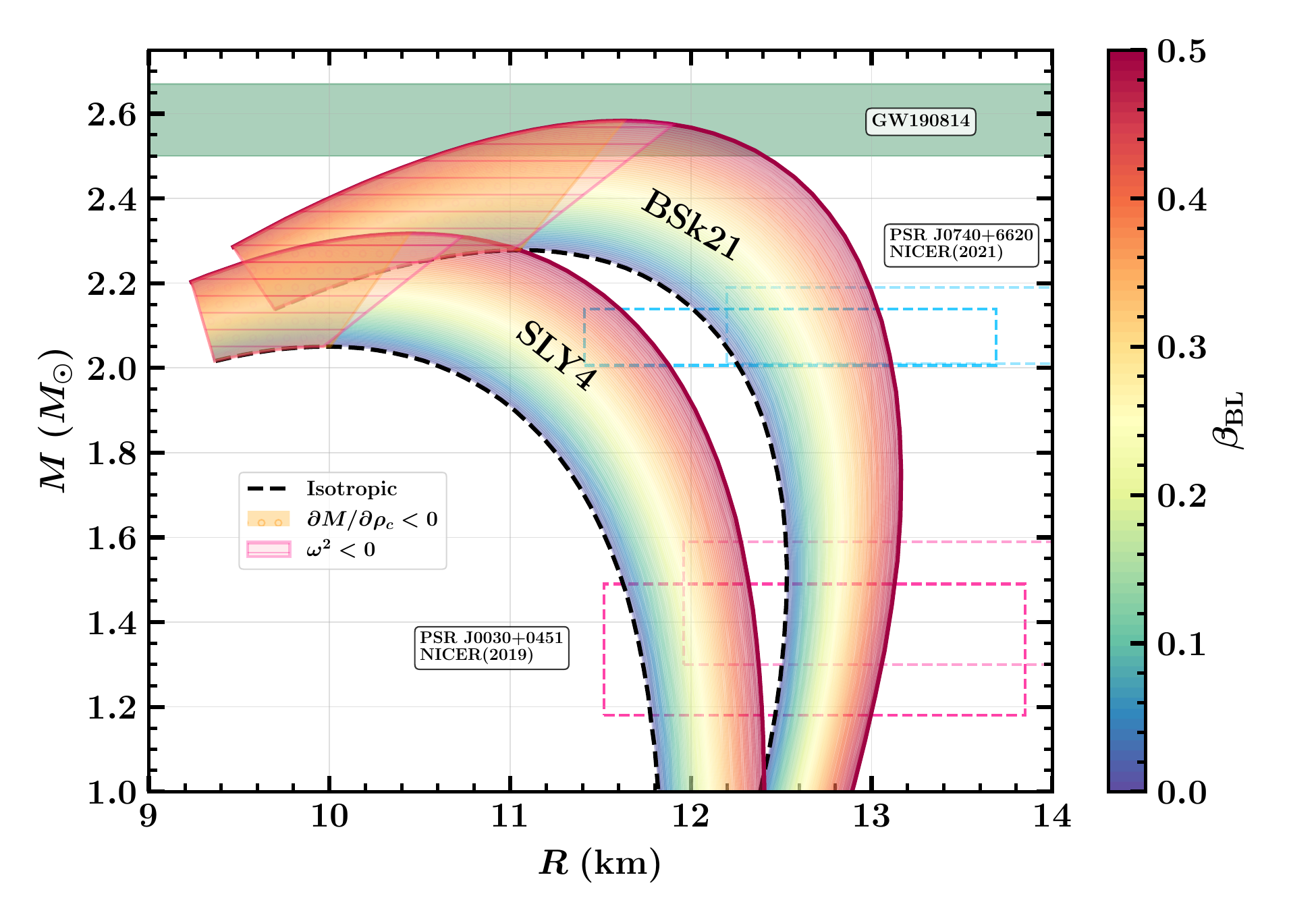}
    \includegraphics[width=0.49\linewidth]{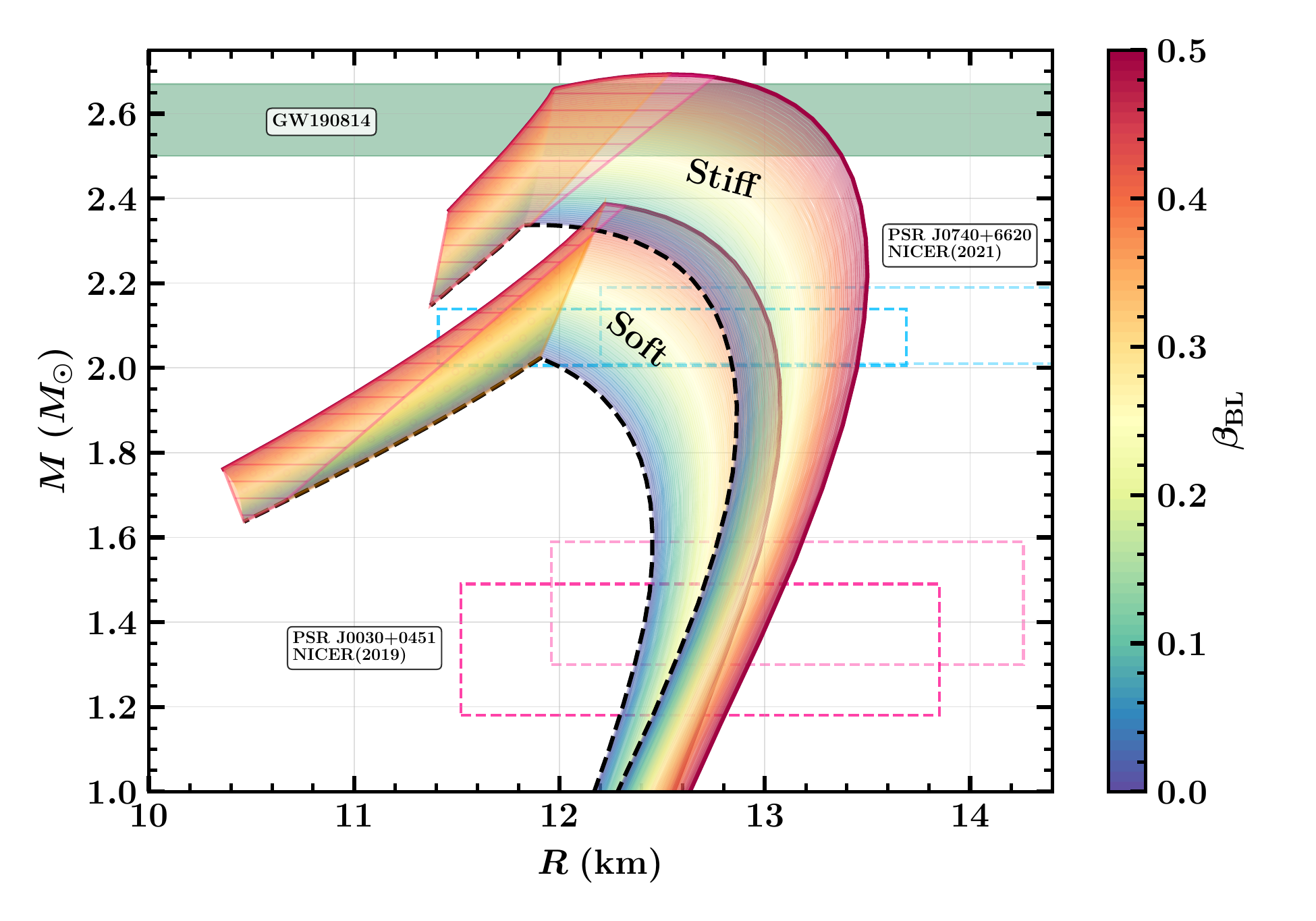}
    \caption{Mass-radius profiles for anisotropic NS with $0 < \beta_{\mathrm{BL}} < 0.5$ corresponding to respective EOSs as depicted in the picture. The lower dark and light pink boxes depict the constraints on mass and radius from the 2019 NICER data of PSR J0030+0451 by Riley and Miller \cite{Miller_2019,Riley_2019}. The upper dark and light blue boxes show the 2021 NICER data with X-ray MultiMirror observations of PSR J0740+6620 \cite{Miller_2021}. Observational mass data from merger event GW190814 is shown as a horizontal green band \cite{GW190814}.}
    \label{fig:MR_Profile}
\end{figure*}

Since we intend to investigate the stability of anisotropic NSs against radial perturbations, it is indeed crucial for a carefull selection of an anisotropic model. Authors in ref. \cite{Pretel2020} have also investigated radial stability of anisotropic NSs for different kind of anisotropic models. One of the key findings in that paper indicates that the star configurations that produces the maximum mass $\left( \left. \frac{\partial M}{\partial \rho_c } \right|_{\bar{\rho}_{c}} = 0 \right)$ align with the onset of instability obtained from the squared frequencies of the fundamental radial mode $\left( \omega^2 (\bar{\rho}_{c}) = 0\right)$ for the Horva et al. \cite{Horvat_2011} and Raposo et al. \cite{Raposo_2019} approaches, but not for the Bowers-Liang \cite{BL-Model} and Herrera-Barreto \cite{Herrera2013} models. Therefore, we opt for the formulation presented by Bowers and Liang \cite{BL-Model} as our anisotropic model of choice. Its stability limit is influenced by the inclusion of anisotropy, a feature that aligns with our specific interest. This model permits a nonzero tangential pressure $P_t$ and introduces an anisotropy parameter $\beta_{ BL}$, which is used to quantify the degree of anisotropy in the system. These models were chosen to provide a comprehensive and realistic description of the properties of NSs within the scope of this study. The anisotropy pressure is defined as 
\begin{equation}
   \sigma = P_{t}-P_{r} = \beta_{\mathrm{BL}} \left(\mathcal{E}+P_{r}\right)\left(\mathcal{E}+3 P_{r}\right) \frac{r^{2}}{e^{-2 \lambda}}. 
   \label{eq:BL}
\end{equation}
The following are some conditions that the anisotropic NS must satisfy \cite{Estevez-Delgado2018,Sulaksono}
\begin{enumerate}
\label{aniso-cond}
  \item Anisotropy shouldn't exist at the center of the NS or $P_{r} = P_{t}$ at $r=0$.
  \item $P_{r}$ and $P_{t}$ must be positive throughout the star. The null energy$(\mathcal{E})$, the dominant energy$(\mathcal{E}+P_{r},$ $ \mathcal{E}+P_{t})$, and the strong energy$\left(\mathcal{E}+P_{r}+2 P_{t}\right)$ must always be positive inside the star.
  \item Speed of sound inside the star must never be negative or exceed the speed of light ($c=1$ taken in this study). Hence the sound speed in radial and tangential direction must hold the following; $0<c_{s,r}^{2},c_{s, t}^{2}<1$.
\end{enumerate}
Before heading on to our main results, lets have a look at the variation of macroscopic properties inside the NS due to inclusion of aniostropy. We have tested all the conditions mentioned above and have found that positive values of $\beta_\mathrm{BL}$ satisfy the aforementioned criteria better than negative values of $\beta_\mathrm{BL}$. Fig. \ref{fig:Sound_speed} shows the speed of sound ($c_{s,t}^{2}=\frac{\partial P_t}{\partial \mathcal{E}}$) inside the NS for the maximum mass configuration of respective EOSs. For negative $\beta_\mathrm{BL}$, $c_{s,t}^{2}$ becomes negative at the surface of the star for all four EOSs, and $c_{s,t}^{2}$ exceeds unity at the center for the hadronic EOSs, violating the anisotropic model conditions. On the other hand, for positive $\beta_\mathrm{BL}$, $c_{s,t}^{2}$ satisfies the causality condition throughout the star \cite{Bhaskar_Biswas,HC_I_LOVE_C}. Therefore, in our further calculations, we will only consider positive anisotropy. We could see a phase transition effect for soft HQPT EOS but not for the stiff one. This is because the central density at which quark-hadron phase transition initiates is reached earlier for soft HQPT EOS before attaining the maximum mass configuration than the stiff ones.

The Mass-Radius (MR) profile for a given EOS can be computed by solving the TOV equation (Eq.(\ref{eq:TOV})) for various central densities, generating a sequence of mass and radius data.  Fig. \ref{fig:MR_Profile} shows the MR profile for the respective EOS. The color gradient describes the anisotropic variation of the MR profile for different values of $\beta_\mathrm{BL}$ ranging from 0 to 0.5. Increasing the value of $\beta_\mathrm{BL}$ results in a higher maximum mass and its corresponding radius for the NS. The orange color patch depicts the unstable region bounded by the maximum mass limit ($\frac{\partial M}{\partial \rho_c}=0$) following the standard criterion of stability, while the pink patch depicts the unstable region bounded by zero eigenfrequency ($\omega = 0$) which is calculated in Sec. \ref{R0}. With an increase in anisotropy, the stability limit determined by zero eigenfrequency ($f$-mode) is attained earlier before reaching maximum mass configuration. This is an obvious result as a NS's stability is decreased due to the inclusion of positive anisotropy. But for soft HQPT EOS, we can see that zero eigenfrequency is attained a lot later after reaching the maximum mass configuration. This happens because, during phase transition, $f$-mode frequency tends to remain constant, which increases the stability of a NS. However the same cannot be observed for stiff EOS, as the zero eigenfrequency is achieved in the hadronic phase (before the phase transition even takes place). We can also see that the maximum mass of the stiff HQPT EOS is attained earlier before the phase transition takes place, while for the soft ones, the maximum mass is attained where the phase transition occurs. Observational data, such as X-ray, NICER, and gravitational wave (GW) data, can be used to constrain the degree of anisotropy within the NS. For example, for BSK21 EOS, values of $0.4<\beta_{BL}<0.5$ satisfy the mass constraint (2.50–2.67 $M_\odot$) of the GW190814 event, while for stiff HQPT EOS, values of $0.2<\beta_{BL}<0.5$ satisfy the mass constraint of the GW190814 event suggesting that one of the merger companions may have been a highly anisotropic NS \cite{Roupas2021}. Moreover, the range of $0.15<\beta_{BL}<0.35$ is consistent with the latest observational data, providing valuable insights into the properties of compact objects in astrophysics.
\section{Tidal Deformability}
\begin{figure*}
    \centering
    \includegraphics[width=0.49\linewidth]{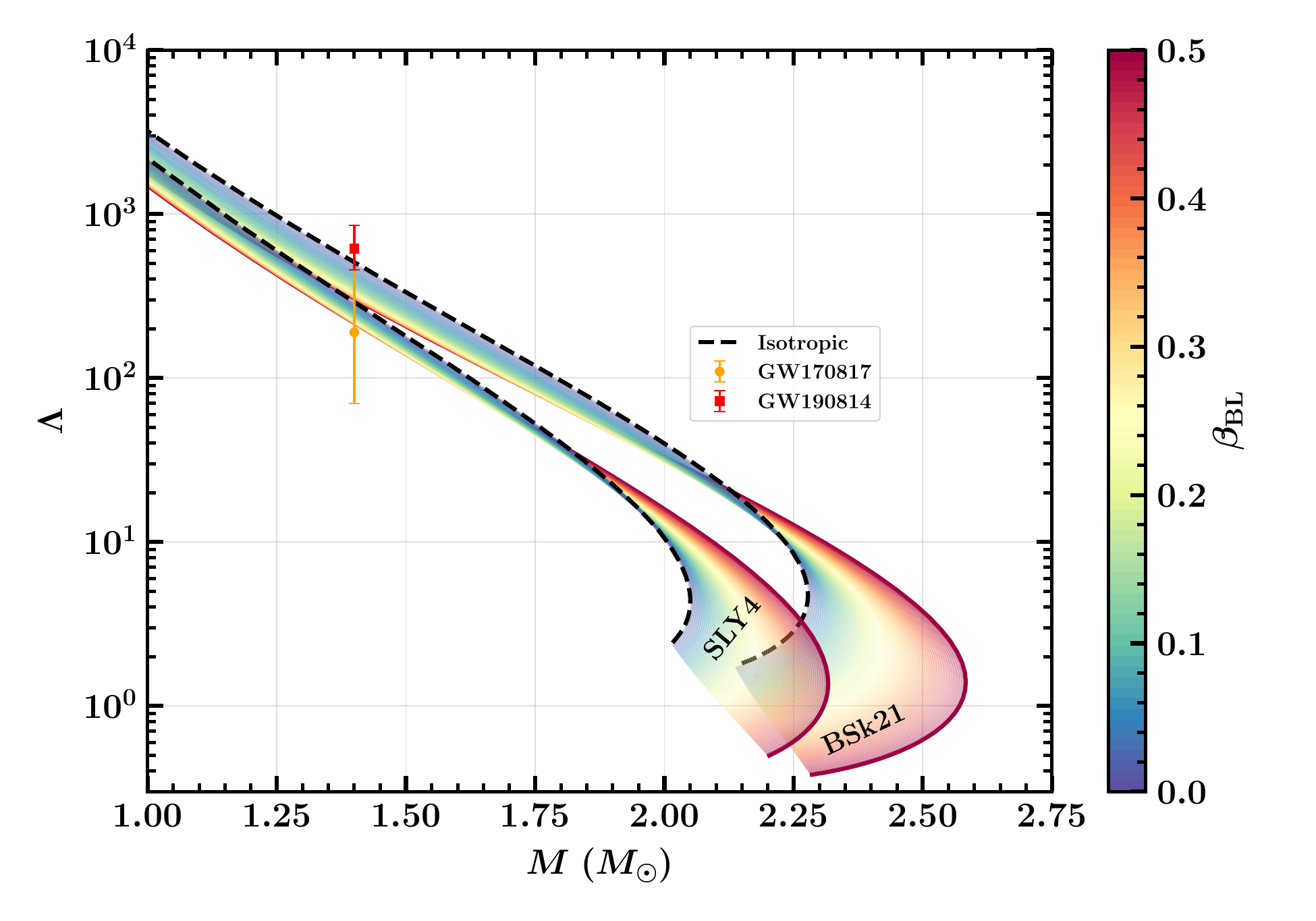}
    \includegraphics[width=0.49\linewidth]{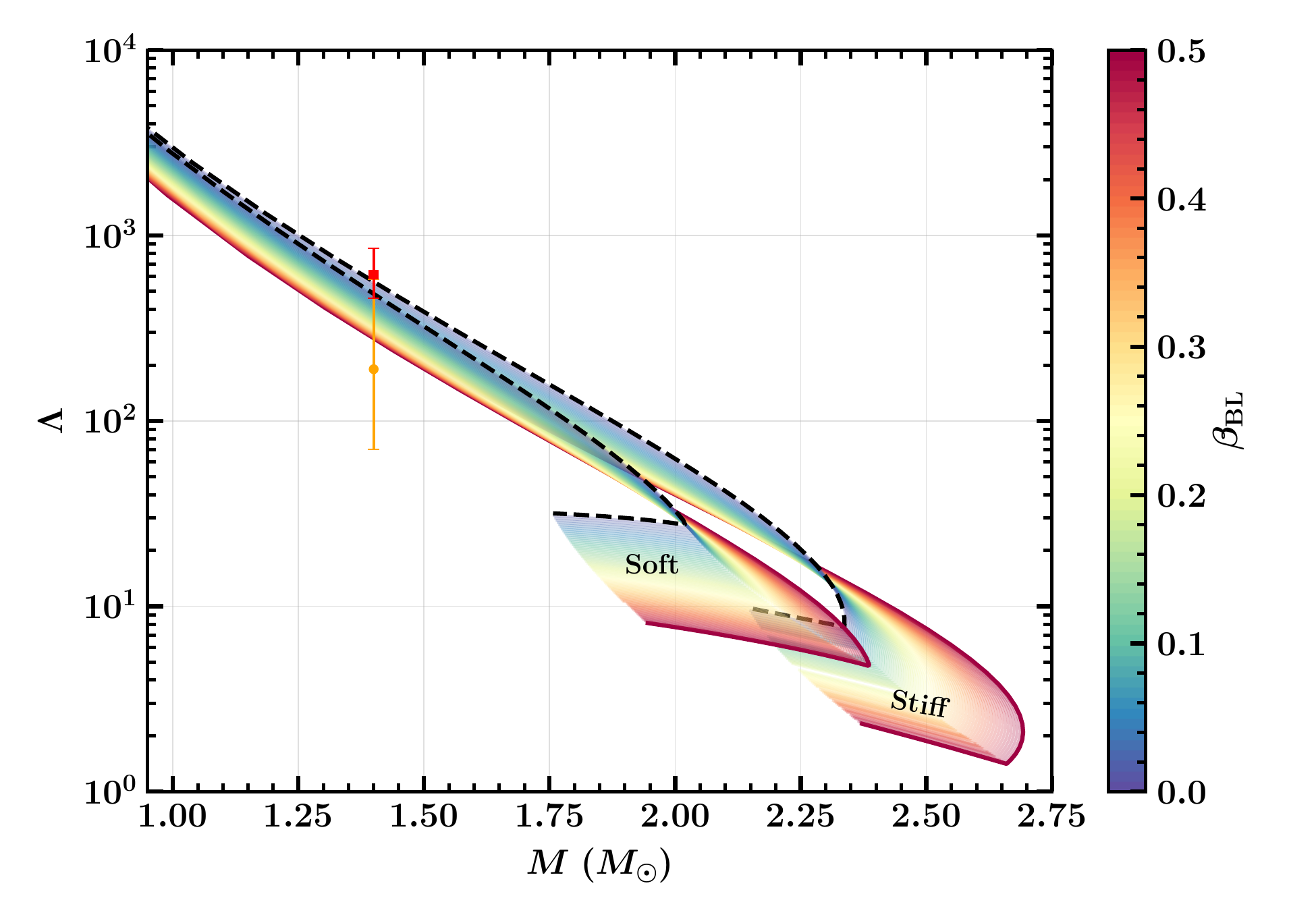}
    \caption{The dimensionless tidal deformability ($\Lambda$) as a function of NS's mass (M) for different $\beta_\mathrm{BL}$ corresponding to respective EOSs as depicted in the picture. LIGO-Virgo's observational constraints for events GW170817 \cite{GW170817} and GW190814 \cite{GW190814} are represented by the error bars.}
    \label{fig:Tidal}
\end{figure*}
NS develops a quadrupole moment $\left(Q_{i j}\right)$ in the presence of an external field $\left(\epsilon_{i j}\right)$ produced by its companion. The quadrupole moment $\left(Q_{i j}\right)$ linearly depends on the tidal field and is given by \cite{Hinderer_2008,Hinderer_2009}
\begin{equation}
Q_{i j}=-\alpha \epsilon_{i j},
\end{equation}
where $\alpha$ is the tidal deformability of a star. $\alpha$ can be defined in terms of tidal Love number $k_{2}$ as $\alpha=\frac{2}{3} k_{2} R^{5}$.

To determine the tidal Love number of an anisotropic NS, we use Thorne and Campolattaro approach \cite{Thorne_and_Campolattaro}, which involves introducing a linear perturbation to the background metric. We obtain the following second-order differential equation for the metric perturbation $H$ from the {$tt$}-component by inserting the fluid and metric perturbations in the perturbed Einstein equation \cite{Bhaskar_Biswas}.
\begin{equation}
\label{eq:tidal}
 H^{\prime \prime}+H^{\prime} \mathrm{U} +H \mathrm{V}  =0 .
\end{equation}
where; 
\begin{align*}
    \mathrm{U} &= \frac{2}{r}+e^{2\lambda}\left(\frac{2 m(r)}{r^2}+4 \pi r(P_r-\mathcal{E})\right) \\
    \mathrm{V} &= 4 \pi e^{2\lambda}\left(4 \mathcal{E}+8 P_r+ \left(\mathcal{E}+P_r\right)\frac{(1+c_{s,r}^2)}{c_{s,t}^2} \right) \\ 
    & \hspace{2cm} -\left( \frac{6 e^{2\lambda}}{r^2} + 4\psi '^2 \right)
\end{align*}
Requiring the regularity of $H$ at the center of the star and the continuity of $H(r)$ and its derivative at the surface yields the value of tidal Love number $(k_2)$ \cite{Hinderer_2008, Damour_Tidal}.
\begin{equation}
\begin{aligned}
k_2= & \frac{8}{5} C^5(1-2 C)^2\left[2\left(y_2-1\right) C-y_2+2\right] \\
& \times\left\{\left[\left(4y_2+4\right) C^4+\left(6 y_2-4\right) C^3-\left(22 y_2-26\right) C^2\right.\right. \\
& \left.+3\left(5 y_2-8\right) C-3\left(y_2-2\right)\right]2 C  +3(1-2 C)^2 \\
& \left.\times\left[2\left(y_2-1\right) C-y_2+2\right] \log (1-2 C)\right\}^{-1}
\end{aligned} 
\end{equation}
where $C$ is the compactness of the anisotropic NS given by $C= {M}/{R}$ and $y_2= {R H^{\prime}(R)}/{H(R)}$. The dimensionless tidal deformability ($\Lambda$) is a useful parameter that can be extracted from GW data \cite{Choi2019} which is defined as follows:
\begin{equation}
    \Lambda= \alpha / M^{5} = \frac{2}{3} k_{2} C^{-5}
\end{equation}
{HQPT EOSs contains a finite energy density discontinuity at the phase transition, for which the term $V$ (coefficient of $H$ in Eq. (\ref{eq:tidal})) involves a singularity $\propto (1+c_{s,r}^2)/c_{s,t}^2$, where $c_{s,t}^{2}=\partial P_t/\partial \mathcal{E}$ and $c_{s,r}^{2}=\partial P_r/\partial \mathcal{E}$ are the respective sound speed squared. As pointed out by authors in Ref. \cite{Postnikov_2010,Han_2019}, in order to capture the delta-function behavior across the point of discontinuity one needs to properly match the solutions of $y(r)= rH^{\prime}(r)/H(r)$ at the point of discontinuity $r_d$;
\begin{equation}
    \left[ y(r) - \frac{4\pi r_d^3 \mathcal{E}(r)}{m_d} \right]_{-}^{+} = 0,
\end{equation}
where $\left[\mathcal{F}(r) \right]_{-}^{+}=\mathcal{F}\left(r_d^{-}\right)-\mathcal{F}\left(r_d^{+}\right)$ and $m_d= m\left(r_d\right)$.} The tidal parameters of anisotropic NSs for all four EOSs used in this analysis are depicted in Fig. \ref{fig:Tidal}. The magnitude of the Love number $k_2$ and its corresponding tidal deformability $\Lambda$ decrease with increasing $\beta_\mathrm{BL}$. The GW170817 \cite{GW170817} event sets a limit of $\Lambda_{1.4}= 190_{-120}^{+390}$, while GW190814 \cite{GW190814} sets a limit of $\Lambda_{1.4}= 616_{-158}^{+273}$ (in the NS-BH scenario). For almost all values of $\beta_\mathrm{BL}$, the predicted value of $\Lambda_{1.4}$ satisfies the GW170817 limit, while $0<\beta_{BL}<0.1$ satisfies the GW190814 limit. However, both $k_2$ and $\Lambda$ start to decrease after crossing {the last} stable configuration.
\section{Slowly Rotating NS and Moment of Inertia}
\begin{figure*}
    \centering
    \includegraphics[width=0.49\linewidth]{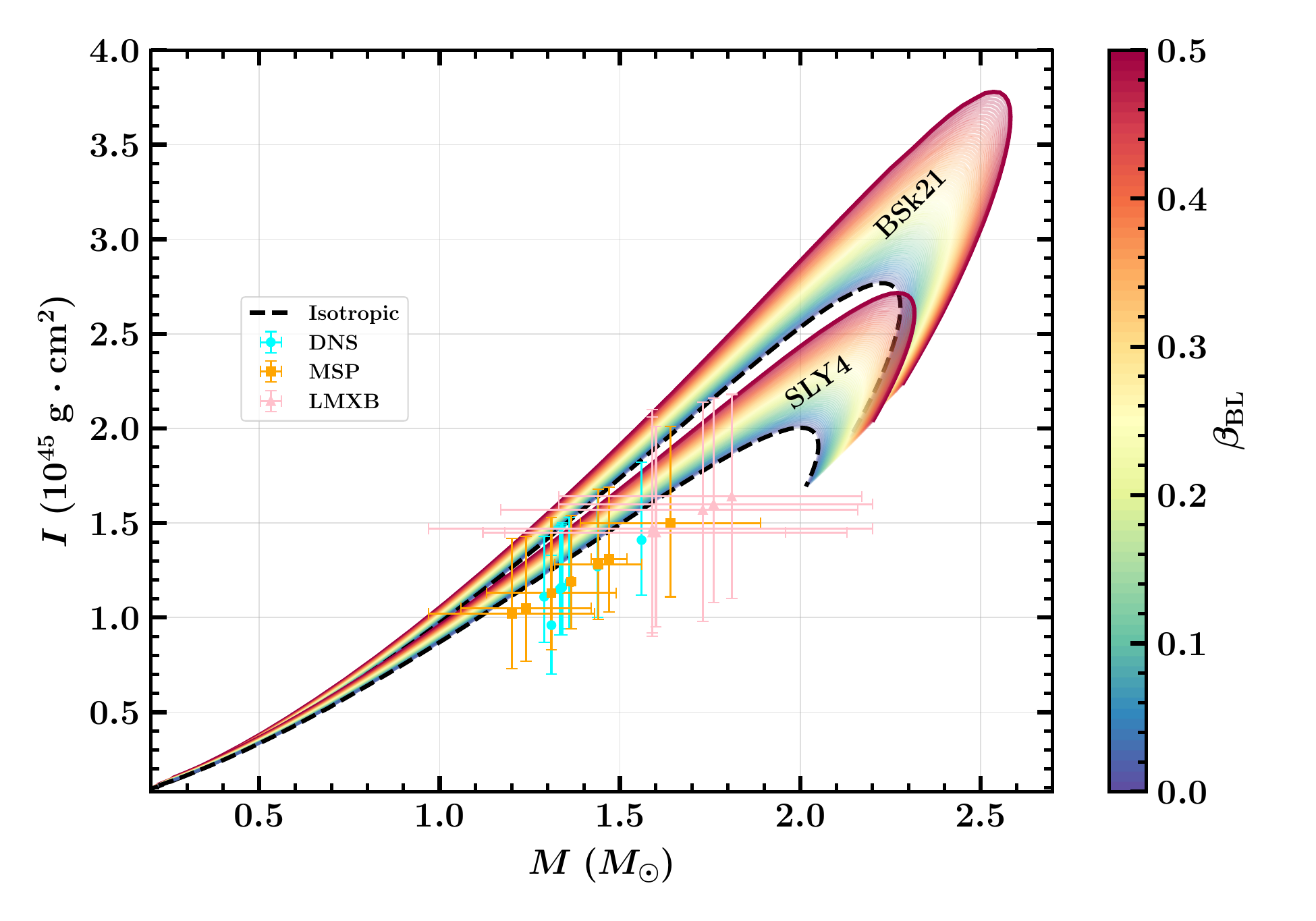}
    \includegraphics[width=0.49\linewidth]{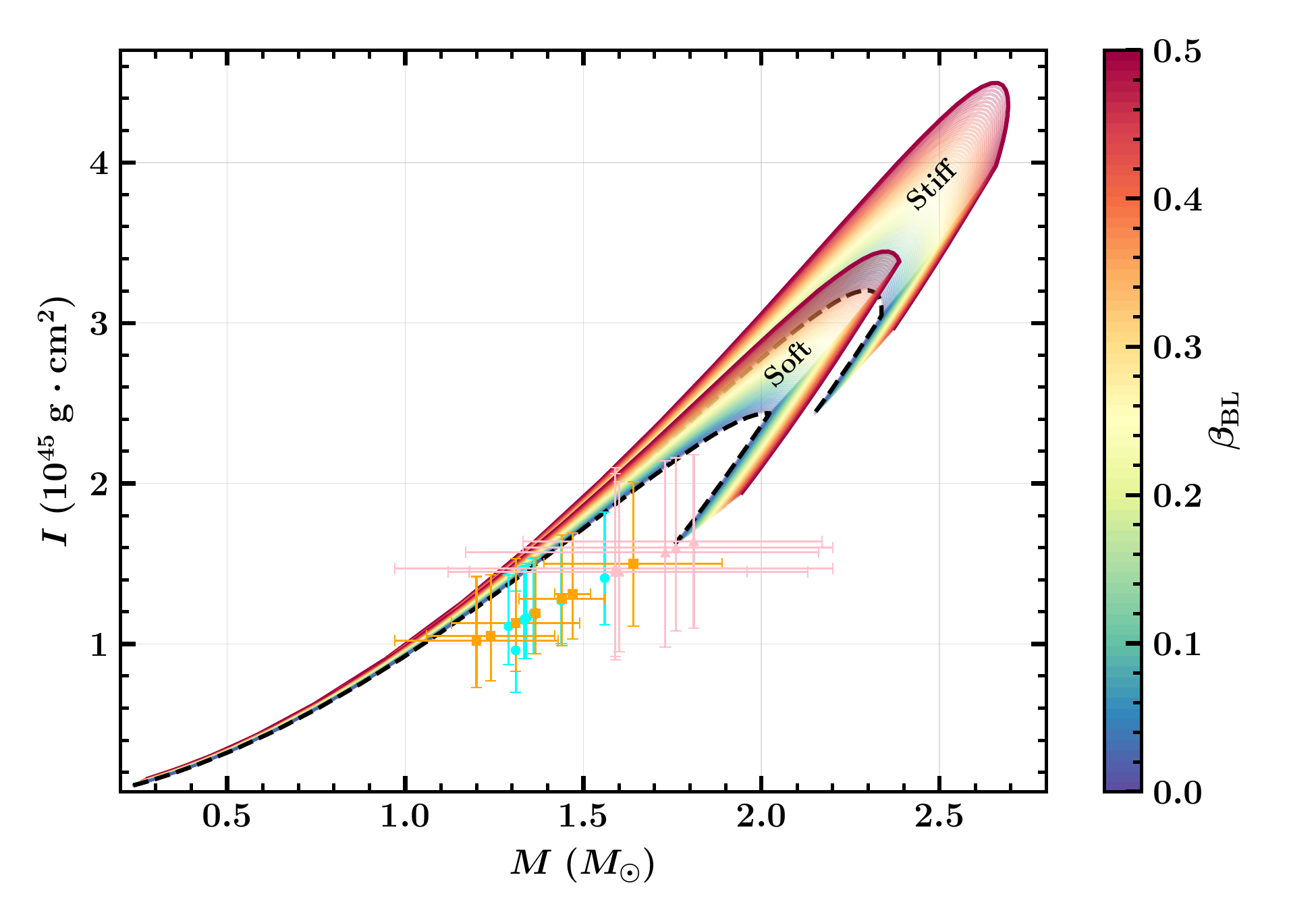}
    \caption{MI as a function of NS's mass (M) with different $\beta_\mathrm{BL}$ for respective EOSs as depicted in the picture. The error bars were calculated based on the results of the several pulsar analyses as done in Ref. \cite{Bharat_and_Landry}.}
    \label{fig:MOI}
\end{figure*}
The line element that describes the space-time inside a slowly rotating anisotropic NS is given by the Hartle-Throne metric as \cite{Hartle-1,Hartle-2,Hartle1973}
\begin{equation}
\begin{aligned}
\label{eq:ht_metric}
d s^{2}=&-e^{2 \psi} d t^{2}+e^{2 \lambda} d r^{2}+r^{2}\left(d \theta^{2}+\sin ^{2} \theta d \phi^{2}\right) \\
&-2 \omega(r) r^{2} \sin ^{2} \theta d t d \phi,
\end{aligned}
\end{equation}
where $\omega(r) = (d \phi / d t)_{\rm ZAMO}$, is the Lense-Thirring angular velocity measured from a zero angular momentum observer (ZAMO) frame of reference, which accounts for the frame-dragging effect. {From the above metric Eq. (\ref{eq:ht_metric}), we can calculate the Einstein-tensor ($G_{\mu \nu}$) in order to solve the Einstein field Eqs. (\ref{eq:einstein}) for a slowly rotating and spherically symmetric sphere whose matter-energy distribution is described by anisotropic stress-energy tensor Eq. (\ref{eq:energy_mom}). We then obtain the following equation from $t\phi$- component of Einstein field equation as described in Ref. \cite{Rahmansyah2020}}
\begin{equation}
\label{eq:ode_mom}
    \frac{1}{r^4} \frac{d}{d r}\left(r^4 J \frac{d\bar{\omega}}{dr} \right)+\frac{4}{r} \frac{d J}{d r}\left(1+\frac{\sigma}{\mathcal{E}+P_r}\right) \bar{\omega}=0,
\end{equation}
where
\[    
J = e^{-\psi}\left(1-\frac{2 m}{r}\right)^{1 / 2},  \quad \text{and} \quad  \bar{\omega}(r) = \Omega-\omega(r) .
\]
By using Eq. (\ref{eq:ode_mom}) and the fact that as $r\rightarrow R$, $J \rightarrow 1$ and $\frac{d \bar{\omega}}{d r} \rightarrow \frac{6 I \Omega}{R^4}$, ($\Omega$, the angular velocity of a uniformly rotating NS and $I$, the moment of inertia (MI)), the MI can be expressed integrally as follows: 
\begin{equation}
\label{eq:mom}
   I=\frac{8 \pi}{3} \int_{0}^R \frac{r^5 J \bar{\omega}}{r-2 m}\frac{\mathcal{E}+P_r}{\Omega}\left[1+\frac{\sigma}{\mathcal{E}+P_r}\right] d r, 
\end{equation}
Fig. \ref{fig:MOI} illustrates the MI profile of NSs as a function of its mass for various degrees of anisotropy, computed for all four EOSs. The MI of a NS increases with increasing mass until a stable configuration is attained, after which it begins to decrease. Furthermore, the NS's mass and MI both increase with increasing $\beta_\mathrm{BL}$. The effects of anisotropy on the MI are more significant for high-mass NSs than for low-mass ones. Kumar and Landry \cite{Bharat_and_Landry} derived MI constraints for various systems, such as double NSs (DNS), millisecond pulsars (MSP), and low-mass X-ray binaries (LMXB), and the error bars depict the range of possible values for these constraints.
\section{Radial Oscillation}
\label{R0}
\begin{figure*}
    \centering
    \includegraphics[width=1.0\linewidth]{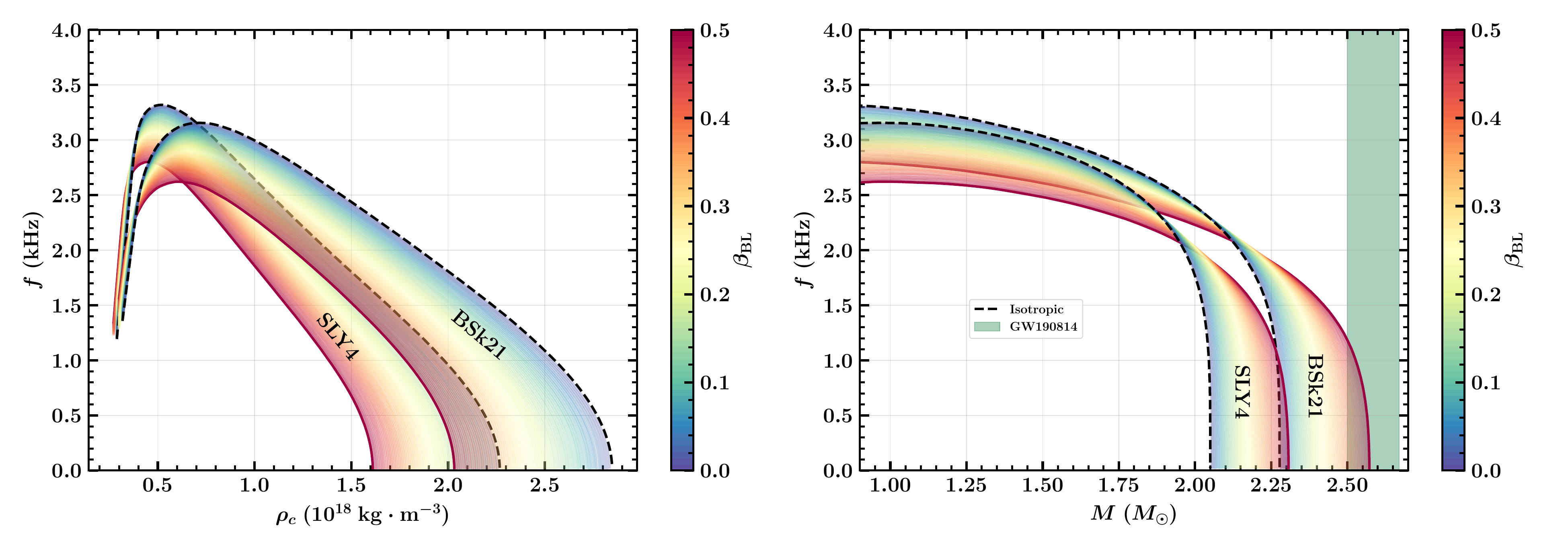}
    \includegraphics[width=1.0\linewidth]{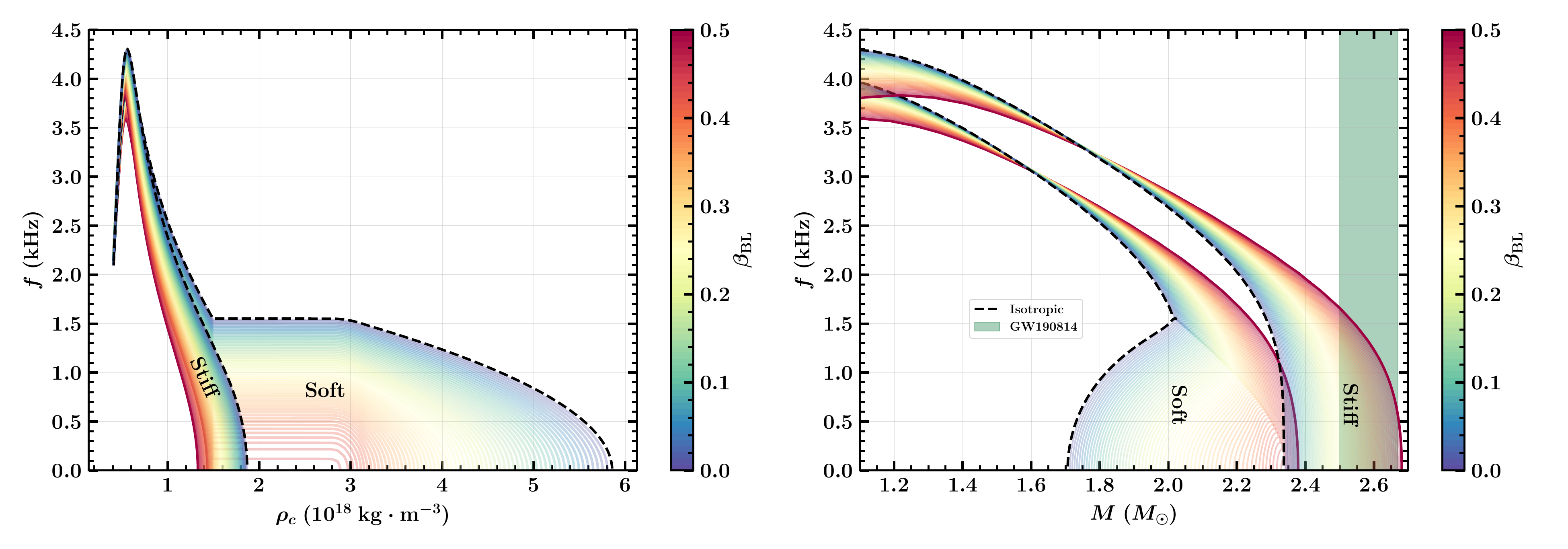}
    \caption{\textit{Left}: $f$-mode frequency as a function of NS's central density ($\rho_c$) with different $\beta_\mathrm{BL}$ for respective EOSs. \textit{Right}: $f$-mode frequency as a function of  NS's mass (M) for different $\beta_\mathrm{BL}$ corresponds to respective EOSs. Observational mass data from merger event GW190814 is shown as a horizontal green band \cite{GW190814}.}
    \label{fig:Radial}
\end{figure*}
To investigate radial oscillation modes, we must examine radial perturbations in an anisotropic fluid configuration. This means that we perturb both fluid and space-time variables without disturbing the spherical symmetry of the background equilibrium configuration. 

Assuming a harmonic time dependence for the radial displacement of a fluid element initially located at $r$ (in the unperturbed condition), which is also known as the Lagrangian displacement, we can write $\delta r = \xi\left(t, r\right)=\xi(r) e^{i \omega t}$, where $\omega$ is the characteristic frequency. By solving the perturbed Einstein equations $\delta G_{\mu \nu}=8 \pi \delta T_{\mu \nu}$, we obtain the linearized radial perturbation equations, which are written as follows \cite{Pretel2020}:
\begin{equation}
\label{eqn:lin_rad_perturb}
\begin{aligned}
& \omega^{2}\left(\mathcal{E}{}+P_{r}\right) e^{2\left(\lambda_{}-\psi_{}\right)} \xi=\Delta P_{r} \frac{d}{d r}\left(2 \psi_{}+\lambda_{}\right)+\frac{d}{d r}\left(\Delta P_{r}\right) \\
& \quad+8 \pi P_{t} e^{2 \lambda_{}}\left(\mathcal{E}{}+P_{r}\right) \xi-\xi\left(\mathcal{E}{}+{P_{r}}\right)\left(\frac{d \psi_{}}{d r}\right)^{2} \\
& \quad+\frac{4 \xi}{r} \frac{d P_{r}}{d r}-\frac{2 \sigma_{} \xi}{r}\left[\frac{d}{d r}\left(2 \psi_{}+\lambda_{}\right)+\frac{4}{r}\right]\\
& \hspace{2cm} -\frac{d}{d r}\left[\frac{2 \sigma_{} \xi}{r}\right]-\frac{2}{r} \delta \sigma \hspace{2cm}
\end{aligned} 
\end{equation}
with perturbations in $P_r, \lambda, \mathcal{E}, \text{and} \ \sigma$ as
\begin{align}
\label{eqn:del_pr}
& \delta P_{r} =-\xi \frac{d P_{r}}{d r}-\gamma P_{r} \frac{e^{\psi_{}}}{r^{2}} \frac{\partial}{\partial r}\left(r^{2} \xi e^{-\psi_{}}\right) +\frac{2}{r} \sigma_{} \xi \frac{\partial P_{r}}{\partial \mathcal{E}}\\ 
& \delta \lambda = -\xi \frac{d}{d r}\left(\psi_{}+\lambda_{}\right)=-4 \pi r\left(\mathcal{E}_{}+P_{r}\right) e^{2 \lambda_{}} \xi \\
& \delta \mathcal{E} =  -\frac{1}{r^{2}} \frac{\partial}{\partial r}\left[r^{2}\left(\mathcal{E}_{}+P_{r}\right) \xi\right]
\end{align}
\begin{equation}
 \begin{aligned}
& \delta \sigma=  2 \beta_{\mathrm{BL}} r^{2} e^{2 \lambda}\left[\left(2 \mathcal{E}+3 P_{r}\right)\left(\Delta P_{r}-r \frac{\xi}{r} P_{r}^{\prime}\right)\right. \\
& \hspace{0.1cm} \left.+\left(\mathcal{E}^2+3 P_{r}^2 + 4 \mathcal{E}P_r \right)(\frac{\xi}{r}+\delta \lambda) +\left(\mathcal{E}+2 P_{r}\right) \delta \mathcal{E}\right] 
\end{aligned}   
\end{equation}
where $\gamma$ is the adiabatic index defined as 
\[
\gamma =\left(1+\frac{\mathcal{E}}{P_{r}}\right) \frac{d P_{r}}{d \mathcal{E}}
\]
Let us redefine the Lagrangian displacement as $\zeta = \xi/r$ in order to rewrite the Eqs. (\ref{eqn:lin_rad_perturb}) and (\ref{eqn:del_pr}) as
\begin{equation}
\label{eqn:CDRPE-1}
\frac{d \zeta}{d r}= -\frac{1}{r}\left(3 \zeta+\frac{\Delta P_{r}}{\gamma P_{r}}+\frac{2 \sigma \zeta}{\mathcal{E}+P_{r}}\right)+\frac{d \psi}{d r} \zeta 
\end{equation}
\begin{equation}
\label{eqn:CDRPE-2}
\begin{aligned}
\frac{d\left(\Delta P_{r}\right)}{d r}= & \zeta\left\{\omega^{2} e^{2(\lambda-\psi)}\left(\mathcal{E}+P_{r}\right) r-4 \frac{d P_{r}}{d r}\right. \\
& -8 \pi\left(\mathcal{E}+P_{r}\right) e^{2 \lambda} r P_{r}+r\left(\mathcal{E}+P_{r}\right)\left(\frac{d \psi}{d r}\right)^{2} \\
& \left.+2 \sigma\left(\frac{4}{r}+\frac{d \psi}{d r}\right)+2 \frac{d \sigma}{d r}\right\}+2 \sigma \frac{d \zeta}{d r} \\
& -\Delta P_{r}\left[\frac{d \psi}{d r}+4\pi \left(\mathcal{E}+P_{r}\right) r e^{2 \lambda}\right]+\frac{2}{r} \delta \sigma
\end{aligned}
\end{equation}
Eq. (\ref{eqn:CDRPE-1}) has a singularity at the origin. Hence the coefficient of $1 / r$ term must vanish as $r \rightarrow 0$.
\[
\Delta P_{r}=-\frac{2 \sigma \zeta}{\mathcal{E}+P_{r}} \gamma P_{r}-3 \gamma \zeta P_{r} \quad as \quad r \rightarrow 0
\]
To maintain the stellar surface boundary condition $P_{r}(R)=0$, the Lagrangian perturbation of the radial pressure must vanish.
\[
\Delta P_{r}=0 \quad as \quad r \rightarrow R
\]
Note that in order to maintain the spherical symmetry configuration of the anisotropic NS, the Lagrangian displacement must vanish at the center, i.e., $\xi(r=0)=0$. However, in our approach, we have already replaced $\xi$ with $\zeta = \xi /r$. Therefore, a natural choice is to take $\zeta=1$ at the center. The differential equations (\ref{eqn:CDRPE-1} and \ref{eqn:CDRPE-2}), subject to the aforementioned boundary conditions, form a Sturm-Liouville eigenvalue problem that can be used to determine the radial oscillation modes with eigenvalues $\omega_{0}^{2}<\omega_{1}^{2} \cdots<\omega_{n}^{2}<\cdots$, where $n$ is the number of nodes inside the anisotropic NS.

\begin{figure}
    \centering
    \includegraphics[trim={0 0.5cm 0 0.5cm},clip,width=0.95\linewidth]{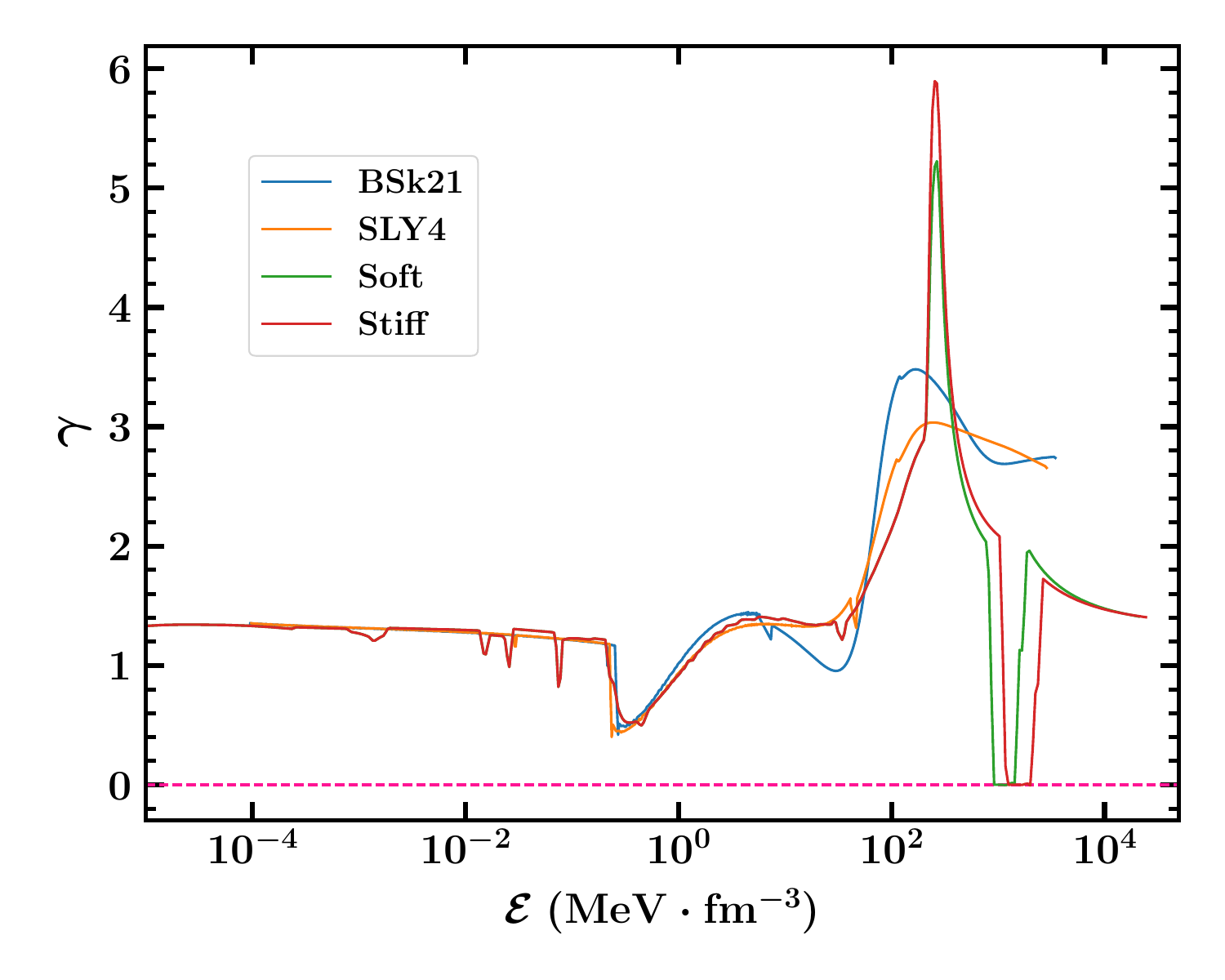}
    \caption{Adiabatic Index ($\gamma$) as a function of energy density ($\mathcal{E}$) for respective EOSs.}
    \label{fig:adiabatic_index}
\end{figure}

\begin{figure*}
    \centering
    \includegraphics[trim={0 2cm 0 2cm},clip,width=\linewidth]{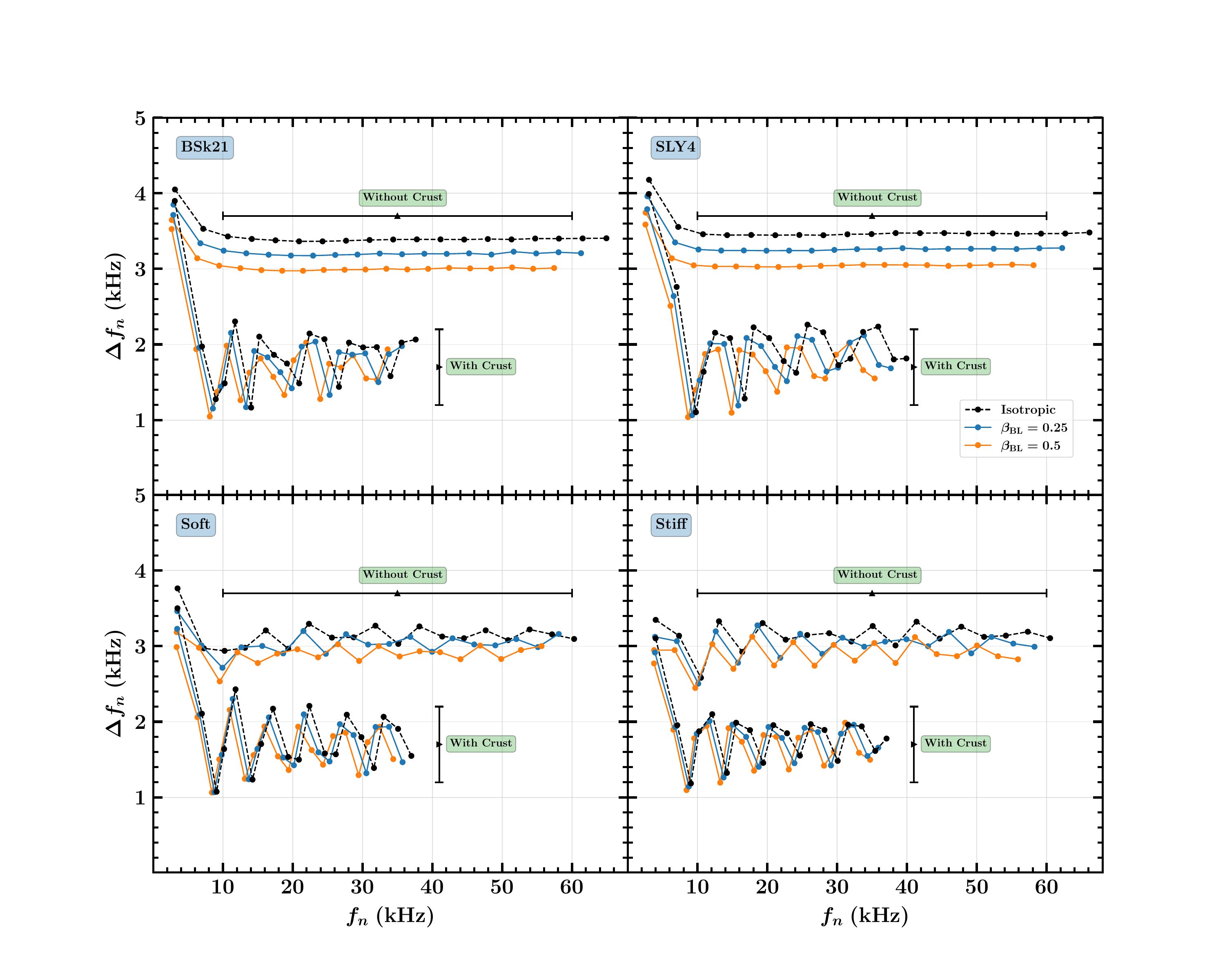}
    \caption{Variation of the large frequency separation with frequency, for the first 20 modes corresponding to canonical mass NS for a system with and without a crust for respective EOSs.}
    \label{fig:delta_f}
\end{figure*}

During calculation of radial modes for HQPT EOSs, junction conditions at the phase transition that relate the values of $\zeta$ and $\Delta P_{r}$ on each side of the transition are needed to determine whether the transition interface oscillates or remains static against radial perturbation. In the vicinity of these interfaces, when subjected to perturbations, two main types of physical behavior arise: either volume elements transform from one phase to another accompanying the fluid movement, or they retain their original state and undergo only stretching or compression. The former is associated with rapid phase transitions, while the latter pertains to slow phase transition \cite{Haensel_1989}. If the time it takes for the process to transform one phase into another is significantly longer than the timescales of the perturbations, it corresponds to slow phase transitions. On the other hand, rapid phase transitions are characterized by conversion rates between phases that occur much faster than the timescales of the perturbations. These conditions were employed in Refs. \cite{Pereira_2018,Rau_2023,Goncalves_2022} to obtain fundamental modes of hybrid stars with slow and rapid conversion. In the case of a slow conversion rate, these conditions are given by
\begin{equation}
\label{eq:slow}
    \left[\Delta P_{r}\right]_{-}^{+}= 0, \quad \left[\zeta (r) \right]_{-}^{+}= 0
\end{equation}
with the $+$ and $-$ corresponding to the high and low density sides of the transition. For a rapid conversion rate, the junction conditions are
\begin{equation}
\label{eq:rapid}
    \left[\Delta P_{r}\right]_{-}^{+}= 0, \quad \left[\zeta (r) - \frac{\Delta P_{r}}{r_d P_r^{\prime}}\right]_{-}^{+}= 0
\end{equation}
One must keep this in mind that the above junction conditions must be employed at the point of discontinuity $(r= r_d)$. As the timescale for the conversion between hadronic and quark phases (the quark matter nucleation timescale) is strongly influenced by the model used \cite{demircik2021dense}, we accommodate both rapid and slow phase transitions due to the lack of precise information about the characteristic timescales of such a phase transition. The radial modes obtained during a rapid phase transition are referred to as $r$- or rapid-modes, while for a slow phase transition, they are denoted as $s$- or slow-modes. 

For the soft HQPT EOS, we found out that, the squared frequency of fundamental $s$-mode as a function of central energy density of NS tends to remain positive after phase transition (suggesting an increament in stability) and supports stable higher-order mass doublets, while $r$-mode suffers a sharp drop and becomes negative (suggesting that the NS loses its stability once its central density crosses phase transition). As for the stiff HQPT EOS, we found out that the zero eigenfrequency is achieved before phase transition takes place. This suggests that the NS loses its stability while being in single-phase state following hadronic part of the stiff HQPT EOS. This happens because the high stiffness of the EOS results in a rapid increase in pressure with density, leading to a higher speed of sound ($c^2 = dP_r/d\mathcal{E}$). Consequently, perturbations travel faster, causing the NS become unstable before the transition energy density is attained. Both $r$-mode and $s$-mode are imaginary in the case of stiff HQPT EOS. Now, the choice of the rate of phase transition to be considered, which is in the interest of this study, becomes somewhat clearer. We must consider the slow phase transition in which hybrid NSs tend to remain stable, allowing us to explore gravitational collapse of hybrid NSs with a dense quark core. More details on the effect of slow and rapid phase transition is discussed in appendix \ref{rs}.

In Fig. \ref{fig:Radial}, we illustrate the relationship between the frequency of $f$-mode oscillations and the central density and mass of anisotropic NSs, considering different EOSs. The $f$-mode instability is identified by the presence of a zero eigenvalue, which occurs at the critical central density $\rho_{\text{critical}}$ \cite{Kokkotas, Ting-Ting}. Once the central density surpasses this critical value, the $f$-mode frequency squared $\omega^2$ becomes negative, resulting in an imaginary frequency. Consequently, the star loses its ability to recover from even minor radial perturbations. Ultimately, the gravitational collapse of the star leads to the formation of a black hole \cite{Harrison, Shapiro}. Furthermore, at lower central densities, a NS can be effectively modeled as a homogeneous, non-relativistic body \cite{Arnett, Vaeth, Shapiro}. In this regime, the eigenfrequency follows the relationship $\omega^2 \propto \rho(4\gamma - 3)$ \cite{Li-strangeon_stars}, indicating that the frequency is predominantly influenced by the density, given that $\gamma$ remains approximately constant at lower densities. Consequently, regardless of the level of anisotropy, the frequency approaches zero when the central density of the star becomes sufficiently low. The outcomes presented in Fig. \ref{fig:Radial} are derived under the assumption of a spherically symmetric background configuration. As anisotropy increases, the $f$-mode frequency decreases. 

For pure hadronic EOSs like SLY4 and BSk21, the $f$-mode curve exhibits a similar trend, with zero eigenfrequency being attained near the maximum mass point. In the case of the soft HQPT-EOS, the stability of the NS is enhanced due to the presence of a quark core. We have considered the phase transition in soft EOS to follow a slow conversion rate. During the phase transition, the $f$-mode frequency remains constant. Consequently, the attainment of the zero eigenfrequency is significantly delayed, occurring after reaching the maximum mass configuration. However, this behavior is not observed for stiff HQPT-EOS which is also considered to undergo a slow phase transition. The maximum mass for the stiff EOS is achieved earlier, prior to the occurrence of the quark-hadron phase transition. Consequently, in the $f$-mode curve for the stiff EOS, no discernible effect of the phase transition is observed. Hence, stiff HQPT-EOS  does not support stable hybrid NSs as the NSs loses stability while still in the hadronic phase.

Fig. \ref{fig:adiabatic_index} displays the behavior of the adiabatic index as a function of density for all the EOSs employed in this analysis. The hadronic EOSs, namely SLY4 and BSk21, exhibit a similar trend in their adiabatic index. However, for the HQPT EOSs, we observe a distinct pattern: the adiabatic index experiences a sudden increase near densities of $300 \ \mathrm{MeV/fm^3}$, followed by a rapid decline, eventually approaching zero during the hadron-quark phase transition. Once the phase transition is completed and the EOS consists solely of quark matter, the adiabatic index continues to steadily decrease with increasing density. Fig. \ref{fig:delta_f} illustrates the variation of the so-called large separation, which is defined as the difference between consecutive nodes, i.e., $\Delta f_n = f_{n+1}-f_n$, with $f_n$ calculated for the canonical mass configuration, for two different degrees of anisotropy. The large separation is a commonly used quantity in asteroseismology and aids in understanding the physics of the star's interior. The behavior of the adiabatic index $\gamma$ in a unified hadronic or HQPT EOS is affected by uneven fluctuations due to the presence of the inner crust. This region is characterized by complex structures collectively known as nuclear pasta, which makes the adiabatic index non-monotonic. The adiabatic index outside the nuclear pasta is about $\gamma=4/3$, determined by the relativistic electron gas, whereas inside the pasta, $\gamma > 2$. For the lowest order mode $(n = 0)$, the role of the crust in radial oscillation is negligible, as it accounts for less than 10\% of the stellar radius and the oscillation nodes are mainly located in the NS core. However, for higher-order modes, some of the oscillation nodes lie in the crust, which is equivalent to lower-order modes of NS without a crust. In other words, to achieve an identical oscillation mode in a NS with a crust and a NS without a crust, the NS with a crust must have additional nodes in its crust. The appearance of a new node in the pasta region results in a peak in $\Delta f$, indicating that the $k^{th}$ peak will have k crustal nodes and (n-k) nodes inside the core, where n is the total number of nodes \cite{galaxies11020060,PhysRevD.107.103039}. Moreover, the points between $k^{th}$ and $(k+1)^{th}$ peaks will also have k and (n-k) nodes in the crust and core, respectively. Therefore, the presence of the crust significantly modulates the eigen-frequency, which results in a squeezed and uneven frequency difference plot.

But when we consider hadronic EOSs without crust, the variation is smooth and consistent with the results reported by Sagun et al. \cite{Sagun_2020}, with the first large separation $\Delta f_0$ being larger than the rest. However, as the degree of anisotropy increases, both $\Delta f_n$ and $f_n$ decrease, but the variation is smooth and consistent, as shown in the figure. This suggests to us that a particular mode in a NS is achieved earlier with the inclusion of anisotropy, resulting in a squeezed large frequency separation plot. Hence, the variation of $\Delta f_n$ and $f_n$ with anisotropy is evidence that the microphysics of the interior of the NS is imprinted in the large separation, a characteristic well-known and also found in main sequence stars, for instance, the Sun. {As stated by authors in ref. \cite{Sagun_2020}, we also found out from our calculations that for a particular anisotropic case, $\Delta f_0$ starts to be a constant proportional to the root mean density of the star ($\sqrt{M/R^3}$) that is independent of $f_n$. Along with it, we also found that $\Delta f_0$ follows a perfect negative correlation with $\beta_\mathrm{BL}$. Hence, we can conclude that $\Delta f_0$ which is the first large separation, is proportional to ($-\beta_\mathrm{BL}\sqrt{M/R^3}$).} Unlikely, when we consider HQPT EOSs without crust, the variation is irregular due to the abrupt change of the adiabatic index at higher energy densities. Even though both $\Delta f_n$ and $f_n$ are higher than systems with the crust which results in a stretched frequency difference plot, the uneven nature of consecutive frequency separation makes it difficult to study the microphysics of the interior of the NS. Yet, we found from our calculations that the first large separation, $\Delta f_0$ has a negative linear correlation with the anisotropic parameter ($\beta_\mathrm{BL}$) and a positive linear correlation with the star's compactness ($C=M/R$). Hence, the macroscopic properties of anisotropic NSs are imprinted in the first large separation for HQPT EOSs. {All the correlations between neutron star parameters with first large frequency difference and anisotropy are listed in appendix \ref{comax}.}

Astroseismology \cite{Hekker2017} is a branch of astrophysics that investigates the dynamics and internal structure of stars by examining their oscillations. These oscillations provide valuable information about the star's properties, such as its mass, radius, age, and chemical composition. While astroseismology is largely employed to investigate the isotropic characteristics of stars, such as their global oscillation modes, it is difficult to directly ascertain anisotropic characteristics. However, some characteristics of star anisotropy can be inferred via indirect techniques. A star's rotation may cause anisotropy. As the star spins, its internal structure becomes deformed, leading to anisotropic effects in the oscillation frequencies. The frequency spacing (Fig.\ref{fig:delta_f}) between the components of these oscillation spectra's frequency multiplets, which represent the rotation-induced splitting, provide information about the star's internal anisotropy and rotation speed. Also, anisotropy can arise from the presence of a magnetic field \cite{10.3389/fspas.2019.00052, Cantiello_2016} within a star. {Magnetic field interactions with oscillations cause mode frequency shifts and amplitude changes. Analyzing these variations offers insights into anisotropic effects.} If in the future we are able to detect radial oscillations, then the signature of $\Delta f_n$ could shed some light on both the anisotropic nature and macroscopic properties of NSs.

The study of SGRBs emission mechanisms provides insights into radial oscillations frequencies, impacting hypermassive NS emissions post-binary NS mergers~\cite{Chirenti_2017}. Distinguishing SGRBs, short hard bursts during mergers, from magneto-emission-driven SGR bursts~\cite{2019PPN....50..613K} is crucial. Furthermore, radial oscillation couples with non-radial modes, amplifying gravitational waves~\cite{passamonti2006coupling, PhysRevD.75.084038}. Present-day GW detectors, including Advanced LIGO, Advanced Virgo, and KAGRA, are projected to achieve a sensitivity of approximately $\sim 2 \times 10^{-22}$--$4 \times 10^{-24}$ strain/$\sqrt{\text{Hz}}$ within the frequency range of 20 Hz to 4 kHz~\cite{PhysRevD.99.102004, abbott2020prospects}. Even third-generation detectors like Cosmic Explorer may attain a sensitivity below $10^{-25}$ strain/$\sqrt{\text{Hz}}$ above a few kHz~\cite{GW170817}, while the underground 10 km long Einstein Telescope is expected to have a sensitivity of $>3\times10^{-25}$ strain/$\sqrt{\text{Hz}}$ at 100 Hz and approximately $\sim 6\times10^{-24}$ strain/$\sqrt{\text{Hz}}$ at around 10 kHz~\cite{Punturo_2010}. Detector sensitivities can be greatly improved by optical adjustments or advanced quantum methods~\cite{PhysRevD.99.102004,2019LRR....22....2D}.

\section{Gravitational Collapse}

In this study, we investigate the dynamical evolution of anisotropic NSs that have become unstable, eventually leading to the formation of an event horizon. The evolution of NSs can be influenced by various instabilities, which can produce gravitational waves \cite{GW170817}. The formation of black holes (BHs) can occur in two scenarios. In the first scenario, a black hole is formed when a massive core collapses, exceeding the maximum mass limit of a NS. In the second scenario, an accreting NS becomes unstable and collapses when it reaches its maximum mass, resulting in the formation of a black hole. The existence of an upper mass limit for NSs is crucial in determining the formation of black holes, whether during their creation or due to the accretion of matter from a companion by mature stars.

The $f$-mode frequency obtained in Sec. \ref{R0} can have two effects on the solution. For real-valued $\omega$ (i.e., $\omega^2 > 0$), the solution is purely oscillatory. However, for imaginary $\omega$ (i.e., $\omega^2 < 0$), the solution grows exponentially, leading to instability in radial oscillations \cite{Horvat_2011}. For anisotropic NSs with a central density $\rho_c$, the fundamental mode $\omega_o$ becomes unstable when the central density exceeds the critical density, i.e., when $\rho_c > \rho_{\rm critical}$. After reaching the critical density configuration, the NS will eventually collapse into a black hole. During the gravitational collapse, massless particles like neutrinos and photons carry thermal energy to the exterior spacetime \cite{PhysRevD.70.084004,PhysRevD.74.024010}, which is a highly dissipative process. We can address this issue using the standard approach of considering a collapsing star's spherical hypersurface $\Sigma$, which separates spacetime into two distinct zones, each characterized by a unique matter-energy distribution. By taking a time-dependent line element \cite{Bogadi2021} to describe the evolution of interior metric and a diffusion approximation in energy-momentum tensor \cite{PhysRevD.70.084004} to describe the energy dissipation during the collapse, we can solve Einstein field equations to get the following (as done in ref \cite{Pretel2020}):
\begin{equation}
\label{sg4}
\mathcal{E} = - \frac{1}{8\pi}\left({\frac{\left(1-e^{2 \lambda}-2 r \lambda^{\prime}\right)}{r^{2} e^{2 \lambda} f}} - \frac{1}{e^{2 \psi}}\frac{3}{4} \frac{\dot{f}^{2}}{f^{2}}\right) ,
\end{equation}
\begin{equation}
\label{sg5}
P_{r} = \frac{1}{8\pi } \left(\frac{1-e^{2 \lambda}+2 r \psi^{\prime}}{e^{2 \lambda} f r^{2}}-\frac{1}{ e^{2 \psi} f }\left(\ddot{f}-\frac{\dot{f}^{2}}{4 f}\right)\right) ,
\end{equation}
\begin{equation}
\label{sg6}
\begin{aligned}
P_{t} = &  \frac{1}{8\pi e^{2 \lambda} f }\left(\psi^{\prime \prime} +\psi^{\prime 2}-\psi^{\prime} \lambda^{\prime}+\frac{\psi^{\prime}}{r}-\frac{\lambda^{\prime}}{r}\right) \\
& \hspace{1cm} -\frac{1}{8\pi e^{2 \psi} f }\left(\ddot{f}-\frac{\dot{f}^{2}}{4 f}\right),
\end{aligned}
\end{equation}
\begin{equation}
\label{sg7}
q  = - \frac{\psi^{\prime}}{8\pi e^{\psi+2 \lambda}} \frac{\dot{f}}{f^2},
\end{equation}
where, the prime and the dot denote differentiation with respect to $r$ and $t$, respectively. $f$ is a time function that depends on $t$ and in the static limit $f\left(t\right) \rightarrow 1$ and we regain Eq. (\ref{eq:metric}), which is the metric for the initially static anisotropic star. The mass constrained by radius $r$ and at time $t$, is expressed as \cite{PhysRevD.70.084004,cahill1970spherical}
\begin{equation}
\label{sg8}
m(t, r) =\frac{r \sqrt{f}}{2 }\left[1-\frac{1}{e^{2 \lambda}}+\frac{r^{2}}{4 e^{2 \psi}} \frac{\dot{f}^{2}}{f}\right] 
\end{equation}

where, it should be mentioned that now, except $\psi=\psi(r), \lambda=\lambda(r)$ and $f=f(t)$ , the energy density, radial pressure, heat flux, anisotropy ansatz, and mass function depend on both $t$ and $r$. The exterior spacetime to $\Sigma$ of the collapsing body is described by the Vaidya metric \cite{Vaidya1951,PhysRevD.90.084011} (for a radiating star). Due to the presence of two distinct spacetime regions with differing geometric properties in a collapsing star, it is necessary to impose junction conditions on the hypersurface $\Sigma$.  As a consequence, the junction conditions signify
\begin{align}
\label{sg16}
\chi_{\Sigma}  =&[r \sqrt{f}]_{\Sigma}=\mathscr{R} \\
\label{sg17}
m_{\Sigma}  =&\frac{ R \sqrt{f}}{2 }\left[1+\frac{r^{2}}{4 e^{2 \psi}} \frac{\dot{f}^{2}}{f}-\frac{1}{e^{2 \lambda}}\right]_{\Sigma} \\
\label{sg18}
z_{\Sigma}  =&\left[\frac{d v}{d \tau}\right]_{\Sigma}-1=\left[\frac{1}{e^{\lambda}}+\frac{r}{2 e^{\psi}} \frac{\dot{f}}{\sqrt{f}}\right]_{\Sigma}^{-1}-1 \\
\label{sg19}
p_{r, \Sigma}  =&\left[q e^{\lambda} \sqrt{f}\right]_{\Sigma}
\end{align}
The non-vanishing radial pressure on the surface of the star leads to an additional differential equation, which enables us to determine the time evolution of the collapse model. Combining Eqs. (\ref{sg5}) and (\ref{sg7}) with Eq. (\ref{sg19}), we obtain
\begin{equation}
\label{sg20}
\frac{d^{2} f}{d t^{2}}-\frac{1}{4 f}\left(\frac{d f}{d t}\right)^{2}-\frac{M}{ R^{2}} \frac{1}{\sqrt{f}} \frac{d f}{d t}=0
\end{equation}
further integrating it we get the following equation
\begin{equation}
\label{sg21}
\frac{d f}{d t}=\frac{4  M}{ R^{2}}\left[\sqrt{f}-f^{1 / 4}\right]
\end{equation}
by applying the static limit we get the integration constant. Then the solution of Eq. (\ref{sg21}) is given by
\begin{equation}
\label{sg22}
t=\frac{ R^{2}}{2  M}\left[\sqrt{f}+2 f^{1 / 4}+2 \ln \left(1-f^{1 / 4}\right)\right].
\end{equation}
At the moment of the formation of an event horizon, the redshift (\ref{sg18}) diverges for an observer who is stationary at infinity. This suggests that the formation of a black hole occurs upon the collapse of an unstable anisotropic star when:
\begin{equation}
\label{sg23}
f_{BH}=\left[\frac{2M}{R}\right]^{4}.
\end{equation}
To model gravitational collapse, we require $0<f \leq 1$, meaning that the temporal function $f$ must monotonically decrease from $f=1$ to $f=f_{BH}$. In other words, time passes from $t=-\infty$ (when the model is static) to $t=t_\mathrm{BH}$ (when the star turns into a black hole), though a temporal shift can be made without losing generality. Using Eq.(\ref{sg17}), we can obtain the mass of the resulting black hole as:
\begin{equation}
\label{sg24}
m_\mathrm{BH}=\frac{2M^{2}}{R}.
\end{equation}

When subjected to radial perturbations, anisotropic NSs in a state of hydrostatic equilibrium undergo oscillations with purely real (fundamental) frequencies. On the other hand, unstable stars, whose lowest oscillation mode exhibits an imaginary frequency, collapse from their equilibrium state and give rise to black holes. In this investigation, we consider that the unstable structures commence in a state of hydrostatic equilibrium before gradually collapsing and eventually forming an event horizon. To study the time evolution of an anisotropic NS during gravitational collapse, we begin by examining an unstable anisotropic NS with an anisotropy parameter $\beta_{BL} = 0.5$ and an eigenfrequency of $\omega^2 = -5 \times 10^7 \ {\mathrm{s^{-2}}}$. We employ the initial configuration obtained by solving the TOV equation (Eq. \ref{eq:TOV}), and subsequently solve equations (\ref{sg4}) to (\ref{sg8}) to determine the time evolution of various NS parameters. The initial configuration of the unstable anisotropic NS, the black-hole time, and the resulting black hole's mass for the EOSs utilized in this analysis are listed in Table \ref{tab:f-Love_coef}.

\begin{figure*}
    \centering
    \includegraphics[width=\linewidth]{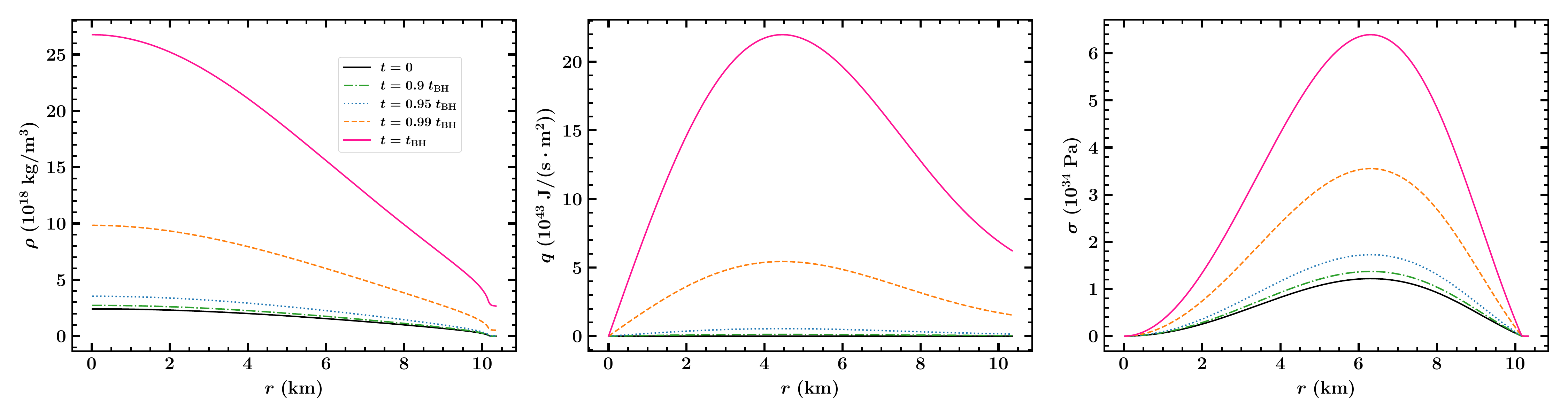}
    \includegraphics[width=\linewidth]{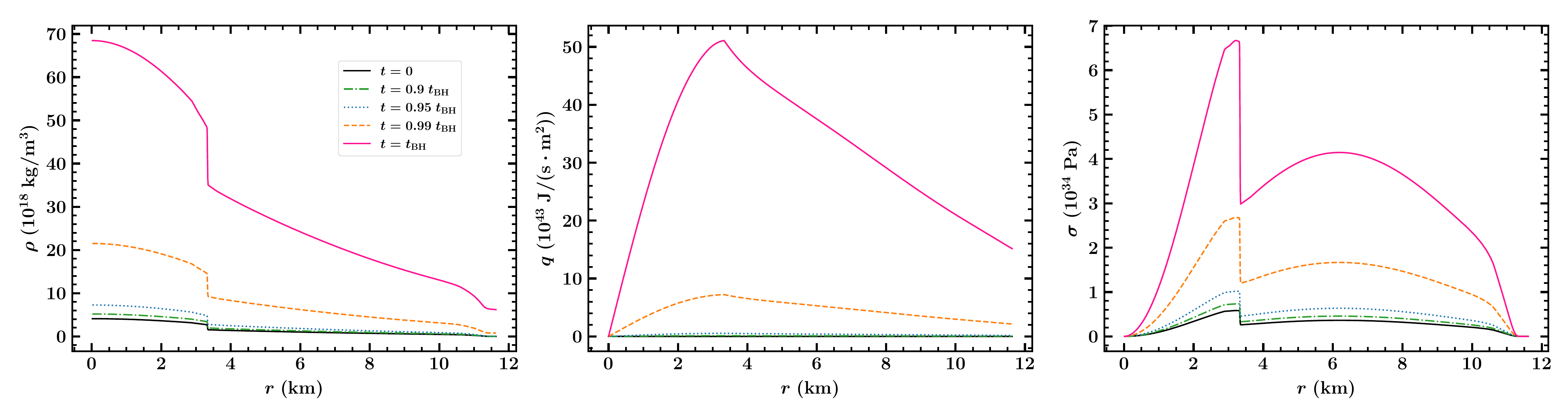}
    \caption{Time evolution of NS's mass density ($\rho$), heat flux ($q$), and anisotropy ($\sigma$) profiles of an unstable anisotropic NS having eigen frequency of $\omega^2 = -5 \times 10^7 \ {\mathrm{s^{-2}}}$ with anisotropy parameter $\beta_{BL} = 0.5$, which initially followed SLY4 EOS (\textit{Upper}) and Soft HQPT EOS (\textit{Lower}).}
    \label{fig:time_evol}
\end{figure*}
\begin{figure}
    \centering
    \includegraphics[width=\linewidth]{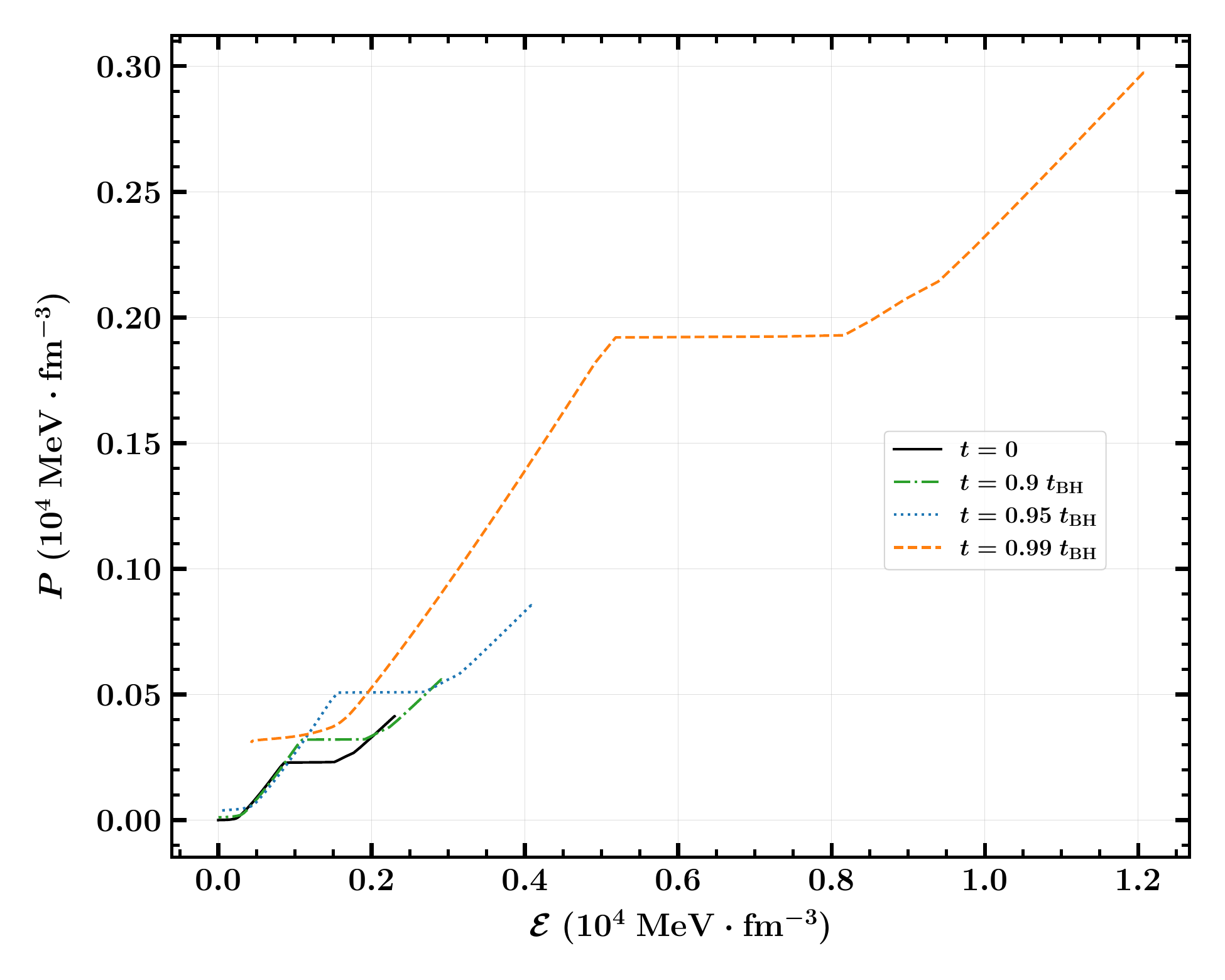}
    \caption{Time evolution of EOS corresponding to an unstable hybrid anisotropic NS having eigen frequency of $\omega^2 = -5 \times 10^7 \ {\mathrm{s^{-2}}}$ with anisotropy parameter $\beta_{BL} = 0.5$. The initially static but unstable NS followed the soft HQPT EOS.}
    \label{fig:GCEOS}
\end{figure}
\begin{figure*}
    \centering
    \includegraphics[trim={0 5cm 0 5cm},clip,width=\linewidth]{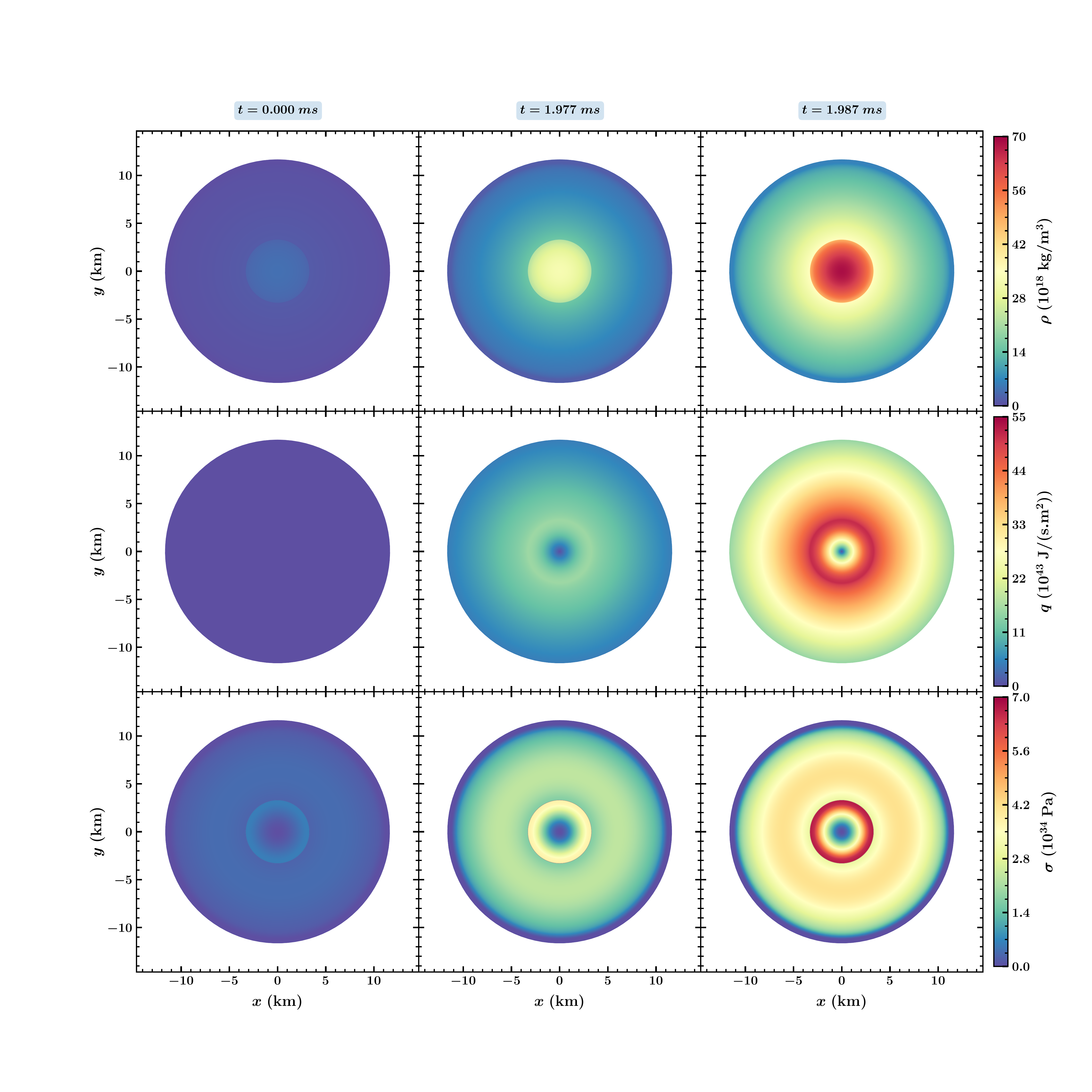}
    \caption{Time evolution of NS's mass density ($\rho$), heat flux ($q$), and anisotropy ($\sigma$) profiles of an unstable anisotropic NS having eigenfrequency of $\omega^2 = -5 \times 10^7 \ {\mathrm{s^{-2}}}$ with anisotropy parameter $\beta_\mathrm{BL} = 0.5$, which initially followed the Soft HQPT EOS.}
    \label{fig:ns-params_time_evolution}
\end{figure*}
\begin{figure*}
    \centering
    \includegraphics[width=0.49\linewidth]{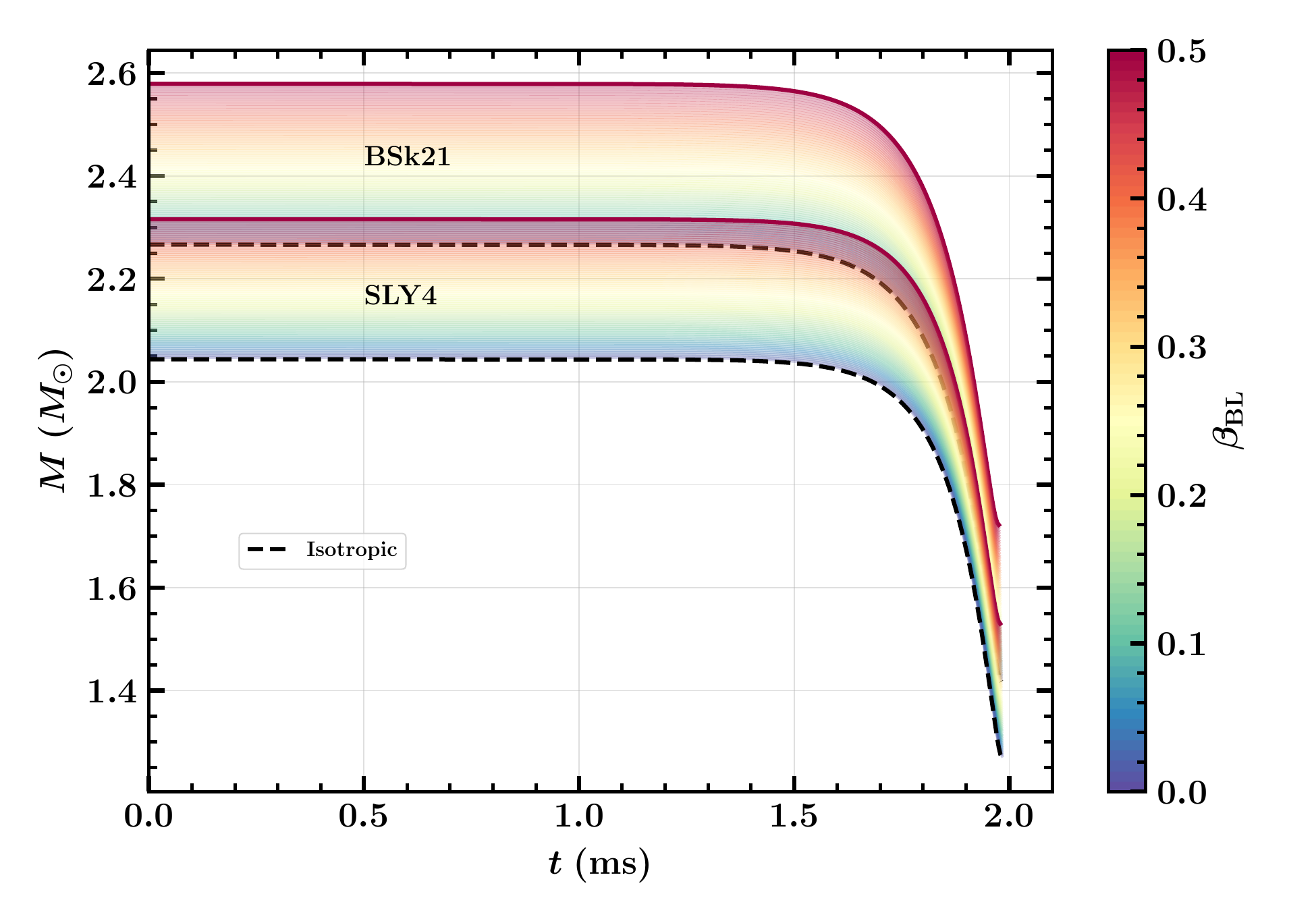}
    \includegraphics[width=0.49\linewidth]{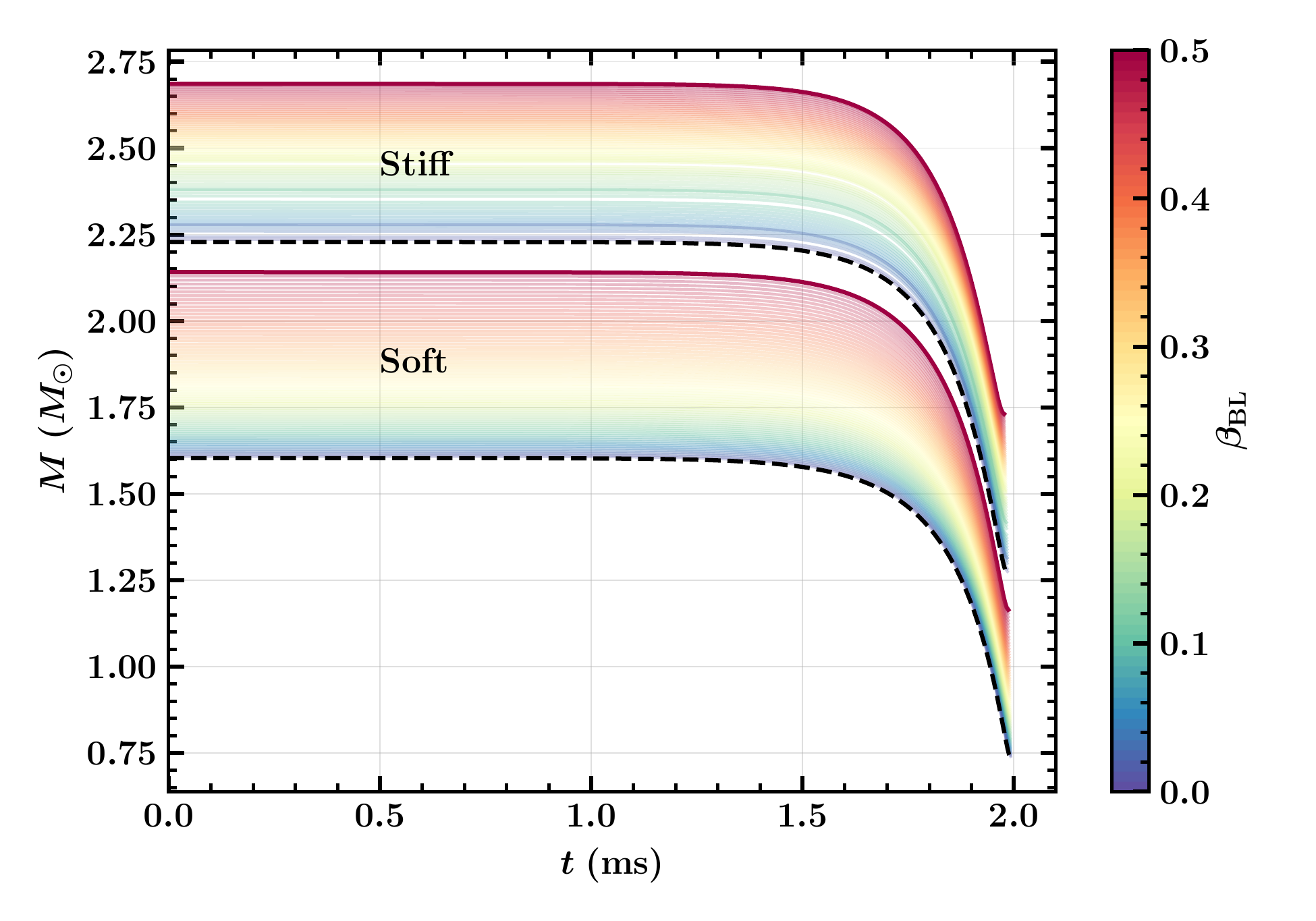}
    \includegraphics[width=0.49\linewidth]{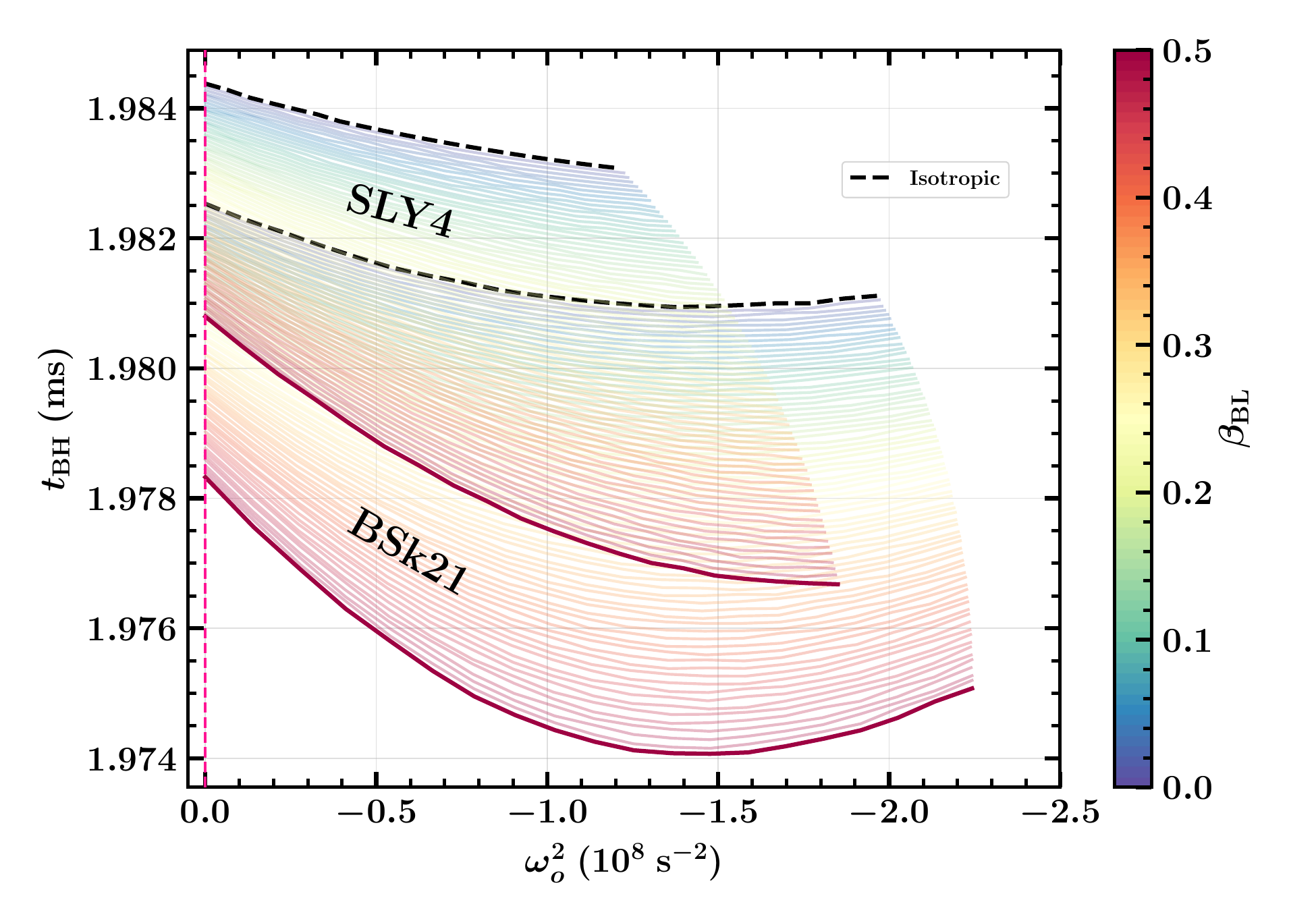}
    \includegraphics[width=0.49\linewidth]{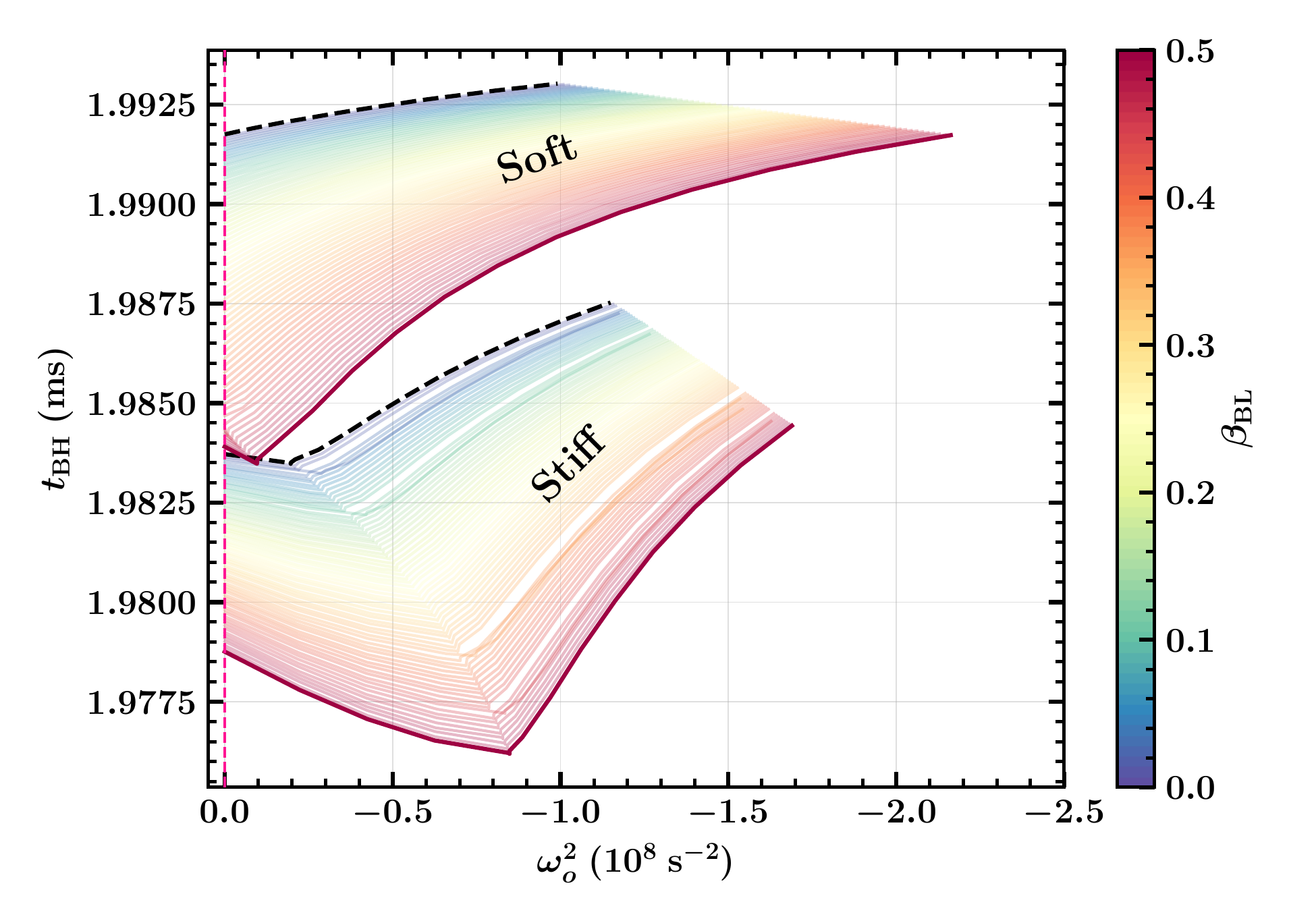}
    \caption{\textit{Top:} Time evolution of the NS's mass for different $\beta_\mathrm{BL}$. \textit{Bottom:} Black-hole time as a function of negative $\omega^2$ for different $\beta_\mathrm{BL}$. The labels in all panels indicate the respective EOS that the initially static but unstable anisotropic NS (at $t=0$) followed.}
    \label{fig:time_evolution_and_w2-tBH_profile}
\end{figure*}
\begin{table}
    \centering
    \caption{This table represents the initial configuration of an unstable NS having an eigenfrequency of $\omega^2 = -5 \times 10^7 \ {\mathrm{s^{-2}}}$ with anisotropy parameter $\beta_{BL} = 0.5$ corresponding to respective EOSs and the time taken and mass of resulting BH.}
    \renewcommand{\arraystretch}{1.4}
    \scalebox{0.9}{
    \begin{tabular}{cccccc}
        \hline \hline
        EOS & $\rho_c \ \mathrm{(10^{18} \ kg/m^3)}$ & $M \ (M_\odot)$ & $R \ \mathrm{(km)}$ & $t_\mathrm{BH} \ \mathrm{(ms)}$ & $m_\mathrm{BH} \ (M_\odot)$ \\
        \hline
        BSk21 & 1.994 & 2.579 & 11.41 & 1.976 & 1.722  \\
        SLY4 & 2.418 & 2.316 & 10.35 & 1.979 & 1.53 \\
        Soft & 4.1 & 2.141 & 11.63 & 1.987 & 1.164  \\
        Stiff & 1.69 & 2.686 & 12.31 & 1.977 & 1.731 \\
        \hline  \hline
    \end{tabular}} \vspace{-3mm}
    \label{tab:f-Love_coef}
\end{table}

Fig. \ref{fig:time_evol} illustrates the time evolution of the internal profile of the aforementioned NS during gravitational collapse. The evolution of mass density ($\rho_c$), heat flux ($q$), and anisotropy ($\sigma$) are presented in the first, second, and third columns, respectively. The profiles as shown in the upper panel correspond to the aforementioned NS, which initially followed SLY4 EOS, while the lower panel's initial NS followed soft HQPT EOS. The Noteworthy changes in the profile were not observed until $t$ exceeded 90\% of $t_\mathrm{BH}$, but significant transformations occurred at $t = 99\% \ t_\mathrm{BH}$. The time-evolution plots obtained for the NS, which initially followed pure hadronic EOS (SLY4), exhibit similarities to those obtained by the authors in reference \cite{Pretel, Pretel2020}. However, this paper presents interesting findings for the evolution of NS, which initially followed an HQPT EOS. The phase transition becomes evident in the lower panel of Fig. \ref{fig:time_evol}. The dominance of quarks in the core is clearly discernible in all three plots. The presence of quarks in the core leads to more abrupt changes in the evolution of mass density and anisotropy compared to the hadronic part. Another interesting result that we concluded is that the transition radius ($r=r_d$) of the collapsing unstable hybrid NS remains unchanged with time. During the accretion and dissipation of matter, the matter has to change its phase while crossing the transition surface. But the quark-hadron transition corresponding to soft HQPT EOS takes place at $\rho=1.8 \times 10^{18} \ \mathrm{kg \cdot m^{-3}}$ (or $\mathcal{E}=1009 \ \mathrm{MeV \cdot fm^{-3}}$), and from Fig. \ref{fig:time_evol} we could see that at the end moment of the collapse the transition takes place at a density 10-20 times higher than the initial phase transition density. This happens because the EOS ($\mathcal{E}=\mathcal{E}(P_r)$ obtained using Eqs. (\ref{sg4}-\ref{sg5})) itself changes during the collapse as shown in Fig. \ref{fig:GCEOS} and hence, the density corresponding to the onset of phase transition also changes with respect to time, thus maintaining the transition radius during the collapse.

During the process of gravitational collapse, both the accretion and dissipation of matter and energy occur. Fig. \ref{fig:ns-params_time_evolution} provides a visual representation of the equatorial cut of the collapsing compact object, which initially followed soft HQPT EOS. The figure depicts the time evolution of mass density ($\rho$) in the first row, heat flux ($q$) in the second row, and anisotropy ($\sigma$) in the third row at different time points: $t = 0 \ \mathrm{ms}$ in the first column, $t = 0.99 \ t_\mathrm{BH}$ in the second column, and $t = t_\mathrm{BH}$ in the third column, within the collapsing NS. The dominance of the quark-bound core is clearly visible in the plots. The central density of the NS undergoes a significant increase due to the accretion of particles from the outer core and crust. The constancy of the transition radius during the collapse is also evident in the evolution of mass density and anisotropy. In the final moments of the collapse, there are substantial fluctuations in mass density and radial pressure, although changes near the surface are minimal. Throughout the dynamical evolution of gravitational collapse, the radial pressure at the surface is no longer zero, unlike in the static case. This is because there exists a radial heat flux according to Eq. (\ref{sg19}). Initially, when the star was in equilibrium, the heat flux was zero throughout the NS. However, towards the end of the collapse, a finite heat flux exists at the surface, while the heat flux is zero at the center. This signifies that the center of the NS at the end of the collapse corresponds to the singularity of a black hole. Next, we investigate the behavior of all unstable anisotropic NSs with an eigenfrequency of $\omega_o^2 = -5 \times 10^7 \ \mathrm{s^{-2}}$. By solving Eq. (\ref{sg8}) at different equally spaced time intervals, we compute the time evolution of the NS's mass ($M$) as shown in Fig. \ref{fig:time_evolution_and_w2-tBH_profile} (upper panel). The labels in Fig. \ref{fig:time_evolution_and_w2-tBH_profile} showcasing the EOS names indicate that the initial NS (at $t=0$) followed the respective EOS. During the process of gravitational collapse, the NS's mass initially remains constant but abruptly decreases near the formation of the event horizon due to the emission of particles into outer spacetime. At the moment of event horizon formation, $m(t_\mathrm{BH},R)$ precisely coincides with the value obtained using the junction condition given by Eq. (\ref{sg24}). The previous calculation of the mass density profile (Fig. \ref{fig:time_evol}), in which the density tremendously increases at the end moments of collapse, suggests that the total mass of the collapsing NS must increase, while Eq. \eqref{sg8} results in a decrease of the NS's mass during the collapse. This is due to the fact that the total mass of the collapsing NS, which is calculated using Eq. \eqref{sg8} is the mass of the incoming particles only, while the mass density ($\rho(r,t)= \mathcal{E}(r,t)/c^2$) which is calculated using Eq. \eqref{sg4} incorporates both incoming and outgoing particles along with the mass equivalent of energy that is involved during the collapse.

The mass of the resulting black hole at the end of the collapse ($m_\mathrm{BH}$) depends on the initial mass ($M$) and radius ($R$) of the NS, allowing us to observe changes in $m_\mathrm{BH}$ due to varying anisotropy. However, we observe minimal changes in the black hole formation time ($t_\mathrm{BH}$). To gain a better understanding of the impact of anisotropy on $t_\mathrm{BH}$, we conduct non-adiabatic dissipative gravitational collapse for all unstable anisotropic NSs with negative $\omega_o^2$ values, comparing the resulting $t_\mathrm{BH}$ with $\omega_o^2$. The outcomes of these calculations are presented in the lower panel of Fig. \ref{fig:time_evolution_and_w2-tBH_profile}, which illustrates how the black hole formation time $t_\mathrm{BH}$ changes with the anisotropic parameter $\beta_\mathrm{BL}$ and the $f$-mode frequency $\omega_o^2$. For the unstable NSs which initially followed pure hadronic EOSs (SLY4 and BSk21), the black hole formation time $t_\mathrm{BH}$ decreases as the anisotropic parameter $\beta_\mathrm{BL}$ increases. This is because increased anisotropy leads to greater instability in the NS, causing it to collapse more rapidly. Similarly, if the central density surpasses the critical density ($\rho_c > \rho_{\mathrm{critical}}$), we observe a decrease in $\omega_o^2$, indicating reduced stability and ultimately resulting in a shorter $t_\mathrm{BH}$. Therefore, we can conclude that $t_\mathrm{BH}$ decreases with decreasing $\omega_o^2$. 

In the case of the unstable NSs which initially followed Soft HQPT EOS, nearly all NSs that exceed the stability limit (or possess imaginary eigenfrequencies) are hybrid stars containing quarks in their core, except for highly anisotropic cases. Remarkably, we observe that the black hole formation time increases with a decrease in $\omega_o^2$. Conversely, for the Stiff EOS, the initial unstable stars are primarily composed of hadronic matter, resulting in a decrease in black hole formation time as expected. However, once the phase transition occurs, we observe a subsequent increase in the black hole formation time. From this, we can indirectly infer that the presence of quarks in the core of unstable anisotropic hybrid NSs can enhance their stability, and in some cases, prevent an unstoppable collapse into a black hole, potentially leading to the emergence of new stationary configurations \cite{Baym_2018}.
\begin{figure*}
    \centering
    \includegraphics[width=0.49\linewidth]{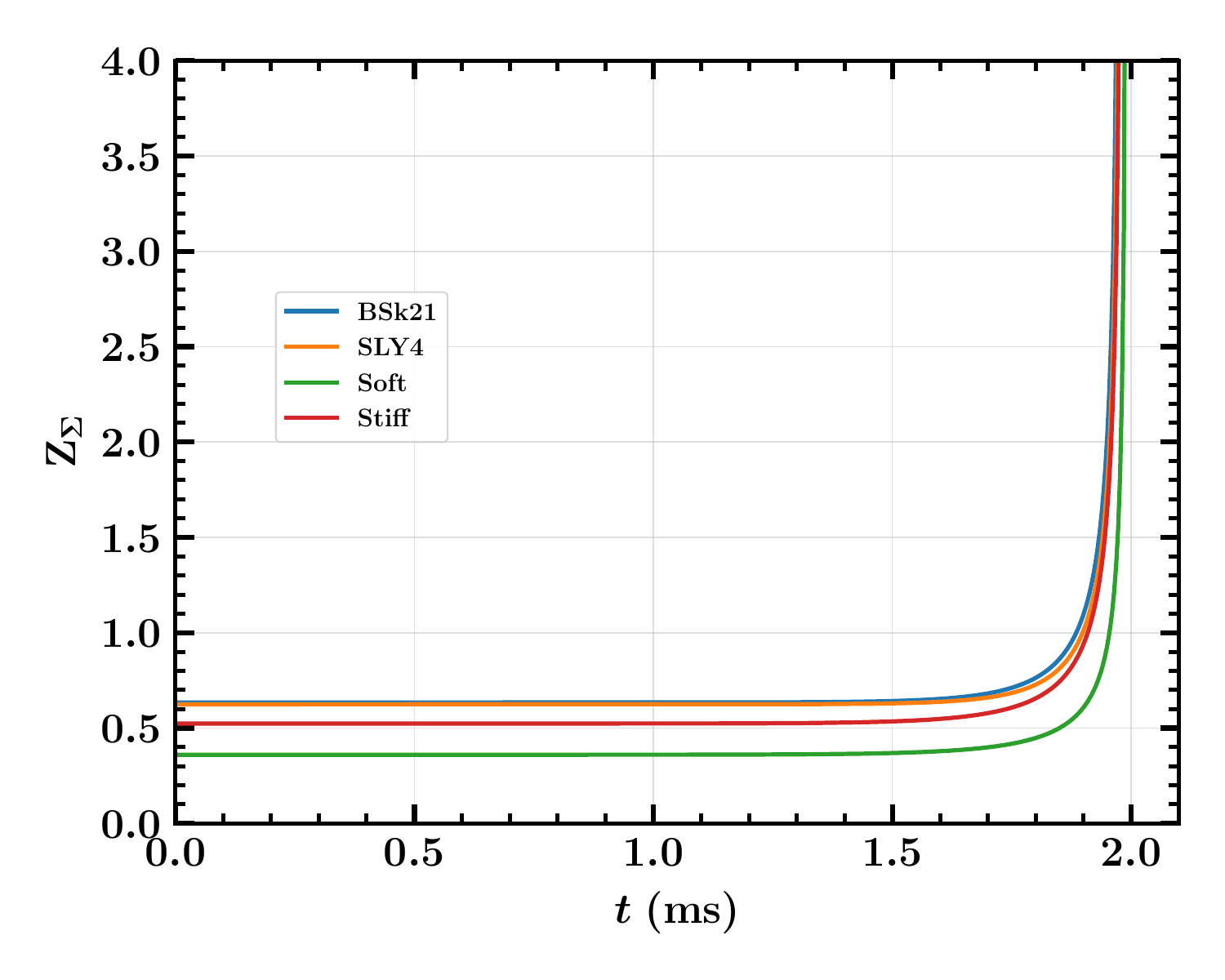}
    \includegraphics[width=0.49\linewidth]{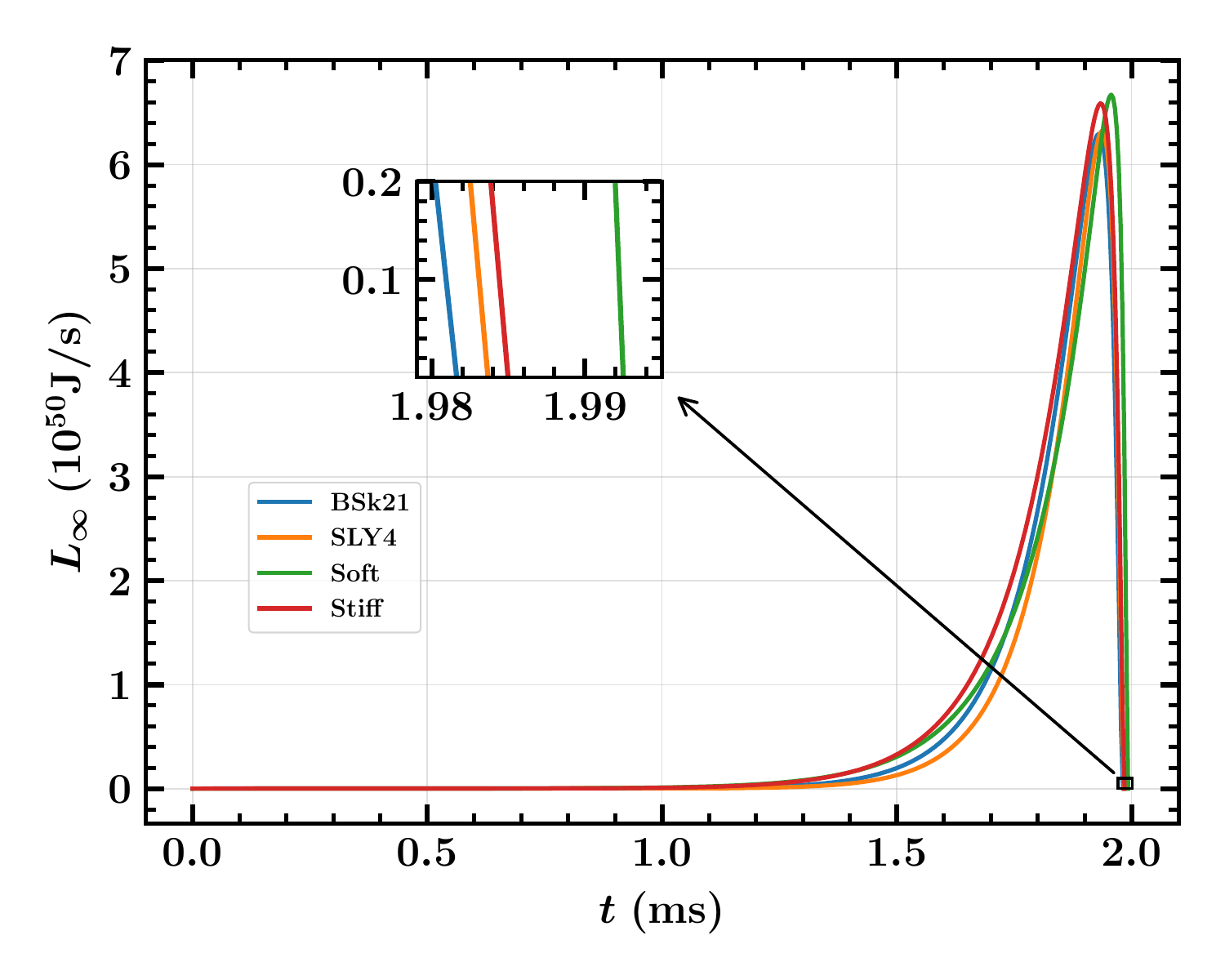}
    \caption{\textit{Left:} Time-Evolution of Gravitational redshift for isotropic NS ($Z_\Sigma \rightarrow \infty$ as $t \rightarrow t_\mathrm{BH}$). \textit{Right:} Time-Evolution of Luminosity for isotropic NS. The labels indicate the respective EOS which the initial unstable isotropic NS followed.}
    \label{fig:Luminosity_and_Red-Shift}
\end{figure*}
\begin{figure*}
    \centering
    \includegraphics[width=0.49\linewidth]{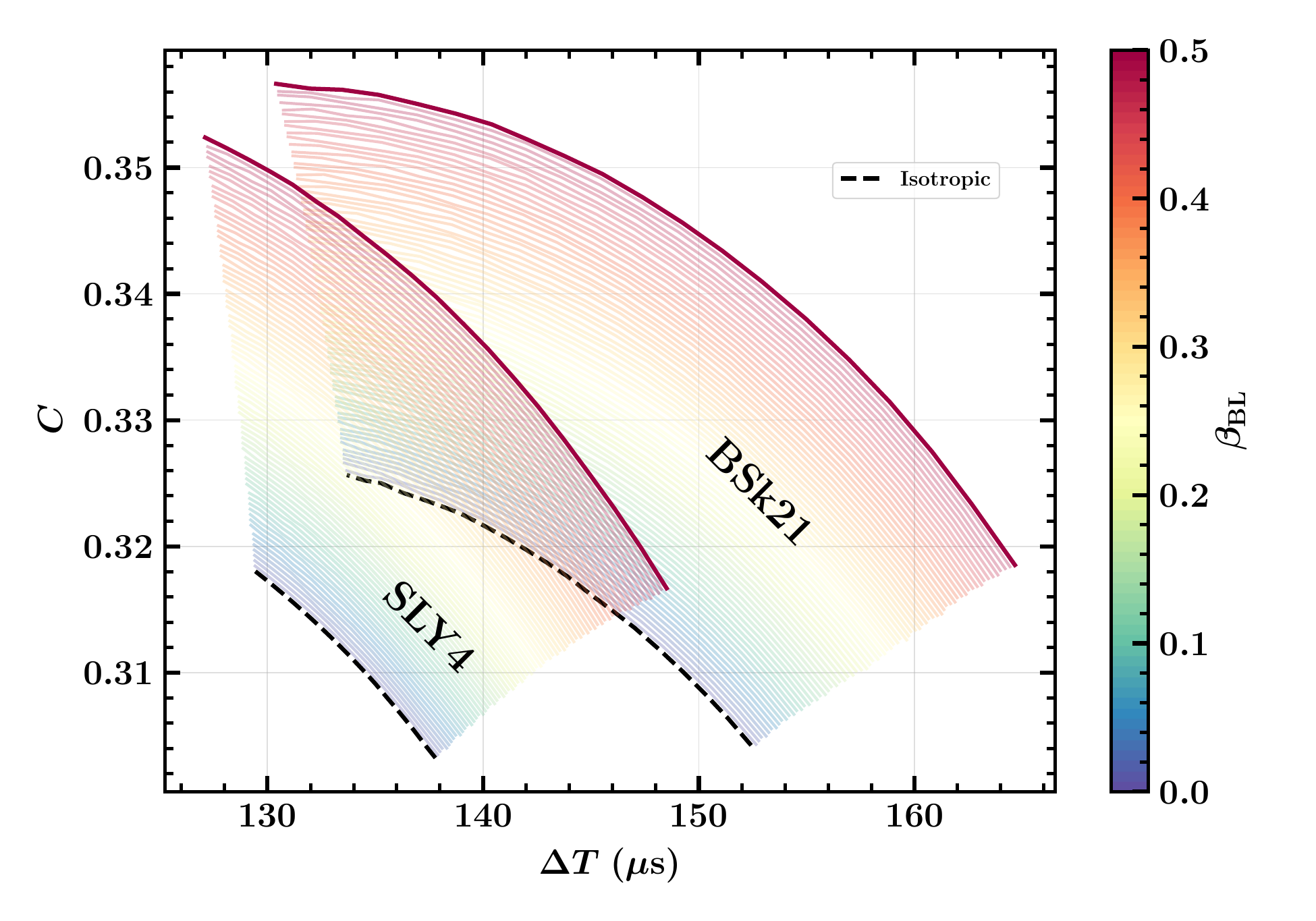}
    \includegraphics[width=0.49\linewidth]{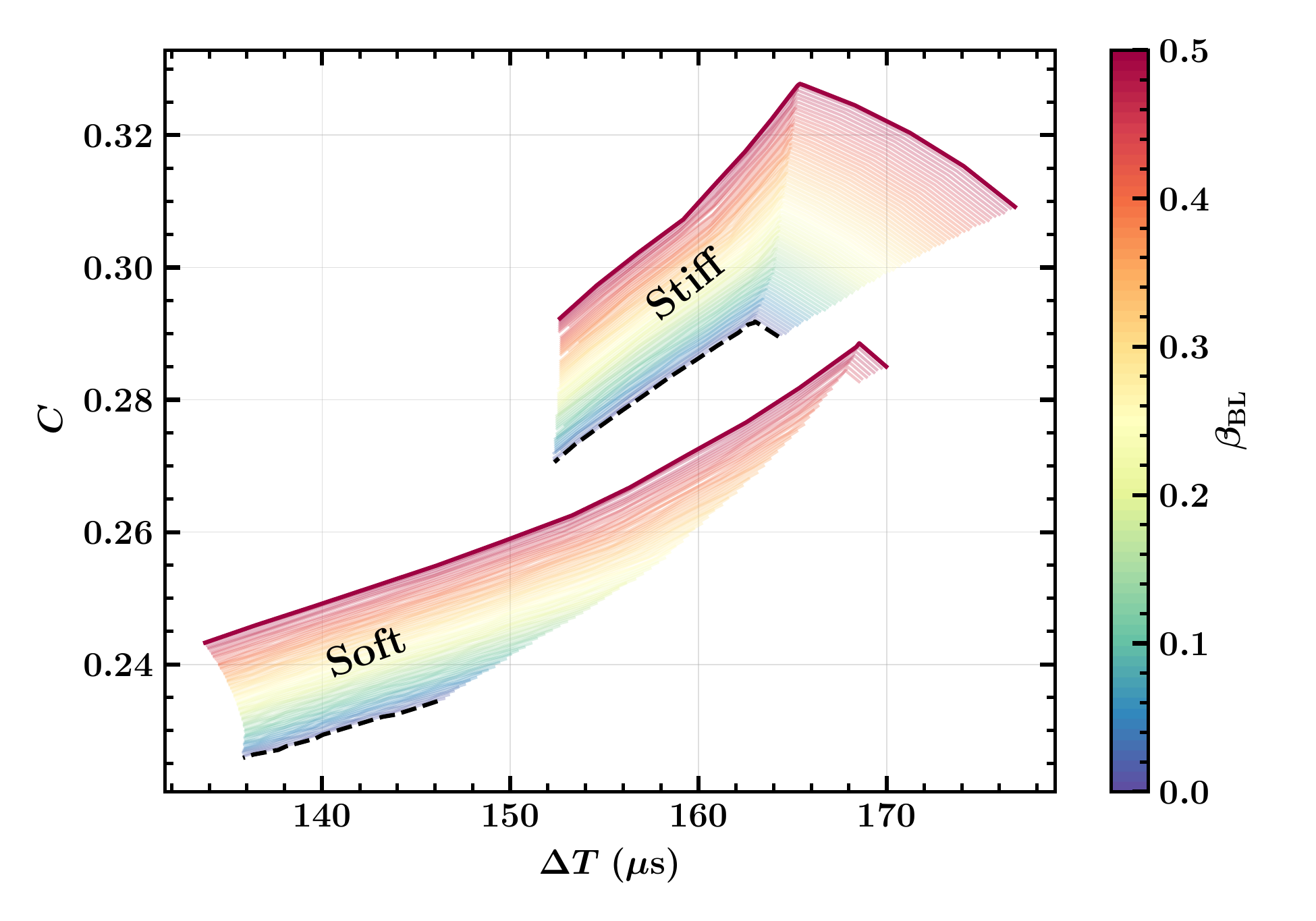}
    \caption{Compactness of static NSs as a function of FWHM (full-width half maxima) $(\Delta T)$ as obtained from luminosity curve corresponding to respective EOSs for different $\beta_\mathrm{BL}$.}
    \label{fig:FWHM}
\end{figure*}

\section{Detectibility}
\label{Redshift}
Detecting the gravitational collapse of unstable NSs can be a challenging task due to the remote locations and extreme physical conditions involved. Nonetheless, researchers employ various methods and observational signatures to study and infer the occurrence of gravitational collapse in NSs. In this section, we present some potential yet challenging ways to detect these highly dissipative events.
Gravitational redshift, also known as the Einstein shift in older literature \cite{EDDINGTON1926}, occurs when electromagnetic waves or photons leaving a gravitational well appear to lose energy. This decrease in energy leads to a decrease in frequency and an increase in the wavelength of the photons. In order to study the time evolution of gravitational redshift for varying anisotropic NSs, we solve Eq. (\ref{sg18}), which describes the surface redshift, at regular intervals of time for all anisotropic NSs with $\omega_o^2 = -5 \times 10^7 \ \mathrm{s^{-2}}$. We also obtain the results for the isotropic case corresponding to the respective EOSs, as shown in Fig. \ref{fig:Luminosity_and_Red-Shift} (left). As expected, the gravitational redshift naturally approaches infinity as the collapse progresses toward the formation of the event horizon. From our calculations, it appears that the anisotropy parameter $\beta_\mathrm{BL}$ does not significantly affect the variation of redshift. This is expected because the redshift is determined primarily by the NS's surface parameters, which are only lightly influenced by anisotropy (since anisotropy vanishes at the surface). Apart from gravitational redshift, variations in luminosity can also be observed during the process of gravitational collapse. As discussed in the previous section, during gravitational collapse, massless particles such as neutrinos and photons carry thermal energy away into the exterior spacetime, resulting in a highly dissipative process. This dissipation of energy can lead to significant changes in luminosity, which can be studied to gain insights into the dynamics of the collapse. To obtain the total luminosity perceived by an observer at rest at infinity, one can refer to the methodology presented in Ref. \cite{G.Pinheiro},

\begin{equation}
    L_\infty = 4\pi f^{3/2}\left[\frac{qr^2}{e^\lambda}\left(\frac{re^\lambda}{2e^\psi} \frac{\dot{f}}{\sqrt{f}} + 1\right)^2 \right]_\Sigma .
\end{equation}

Fig. \ref{fig:Luminosity_and_Red-Shift} (right) illustrates the time evolution of the luminosity ($L_\infty$) of a collapsing NS perceived by an observer at rest at infinity. The initial NS was isotropic in nature and the labels indicate the respective EOS which the initial NS followed. The luminosity exhibits a maximum growth phase, followed by an abrupt decline until black hole formation. These sudden changes in luminosity can be easily detected using photometry techniques \cite{Stetson2013}. The variations in luminosity are primarily caused by the dissipation of energy carried by massless particles during the process of gravitational collapse, as discussed in the previous paragraph. Similar to the surface redshift, the time evolution of luminosity is minimally affected by changes in the anisotropy parameter ($\beta_\mathrm{BL}$). The changes in luminosity primarily depend on the properties of the collapsing object and the dissipative processes occurring within it. To differentiate the black hole formation times for the respective EOSs, a subplot zooming in on the late stages of the gravitational collapse process, where the luminosity approaches zero, is included. It is evident that the NS which initally followed Soft HQPT EOS exhibits the longest collapse time compared to the other EOSs under consideration. This indicates that the specific properties of the EOS, including the presence of quark matter, can significantly influence the duration of the collapse process.

In the study of gravitational collapse of unstable NSs, the exact initiation time of the collapse, or the `BH-time', cannot be measured accurately due to the difficulty of fixing $t=0$. In the absence of being able to measure the BH formation time directly, astronomers and astrophysicists have devised alternative methods to gain insights into the collapse dynamics of compact objects. However, there are methods to gain insights into the collapse dynamics. One such method involves analyzing the luminosity emitted during the collapse process. As material falls onto the forming compact object, it undergoes compression and heating, leading to the emission of radiation across various wavelengths. One important parameter used to characterize the collapse timescale is the full-width half-maxima (FWHM) of the luminosity plot. It represents the time difference in the luminosity curve when the luminosity reaches half of its maximum value. By capturing a bump in luminosity caused by the collapse and measuring the FWHM, valuable information about the collapse process can be obtained. In Fig. \ref{fig:FWHM}, we present the relationship between the compactness of static anisotropic NSs and the FWHM ($\Delta T$) for a sequence of unstable stars corresponding to their respective EOSs. For NSs described by hadronic EOSs, a decrease in compactness leads to an increase in $\Delta T$. Additionally, for a fixed compactness value, an increasing $\beta_\mathrm{BL}$ also increases $\Delta T$. On the other hand, NSs described by hadronic EOSs with a phase transition to a quark core exhibit a linear relationship between compactness and $\Delta T$. As the compactness of these NSs increases, their $\Delta T$ also increases. Notably, the hadronic part of HQPT EOSs behaves similarly to normal hadronic NSs. The establishment of universal relations between $\Delta T$ and other essential parameters of static NSs, such as Mass, Radius, Tidal Deformability, etc., holds the potential to provide valuable physical insights into these collapsing objects. However, it is essential to consider that including phase transition EOSs could disrupt these universal relations \cite{Sandoval_2023}.

\section{Conclusion}
\begin{figure*}
    \centering
    \includegraphics[width=0.5\linewidth]{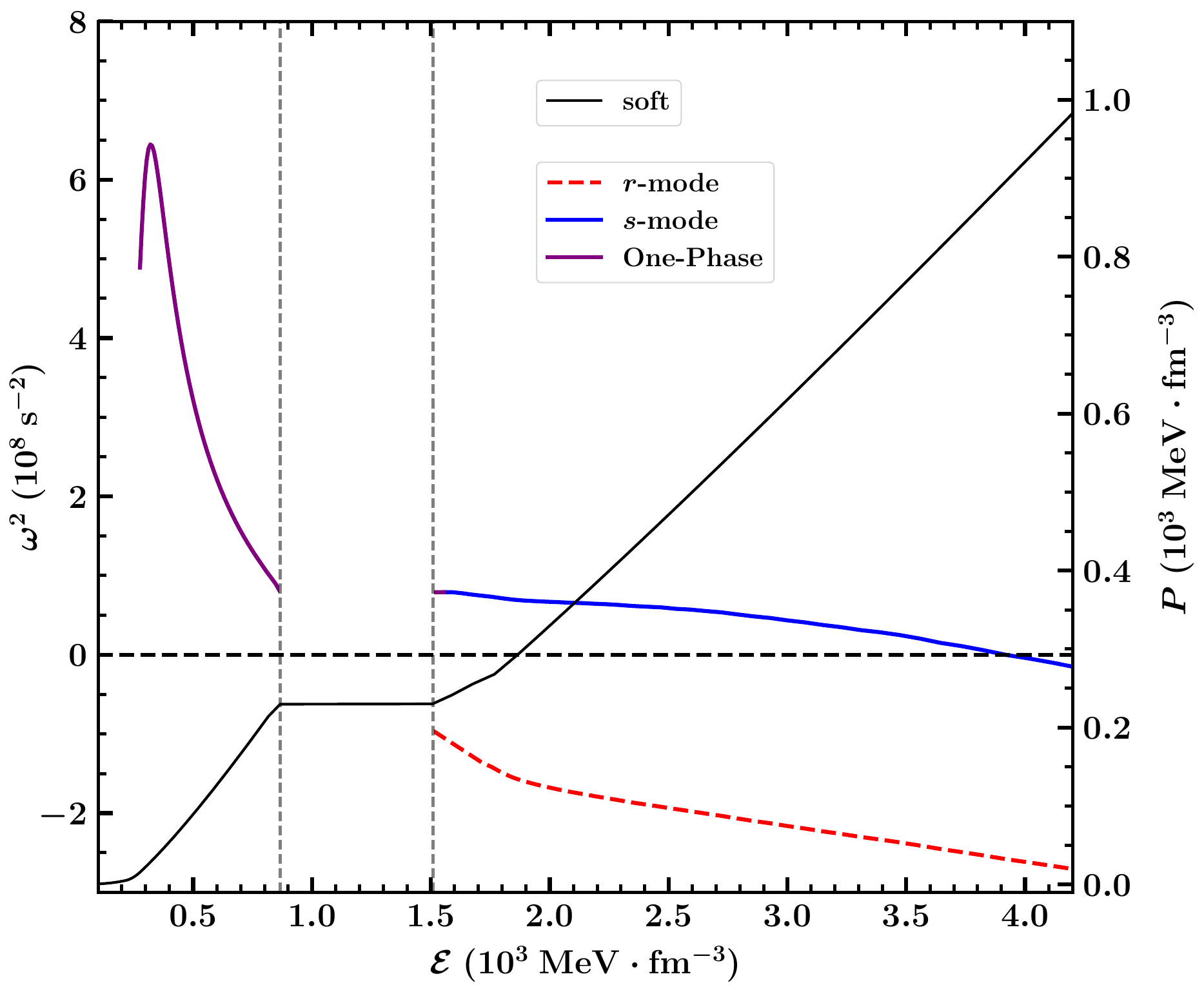}
    \includegraphics[width=0.5\linewidth]{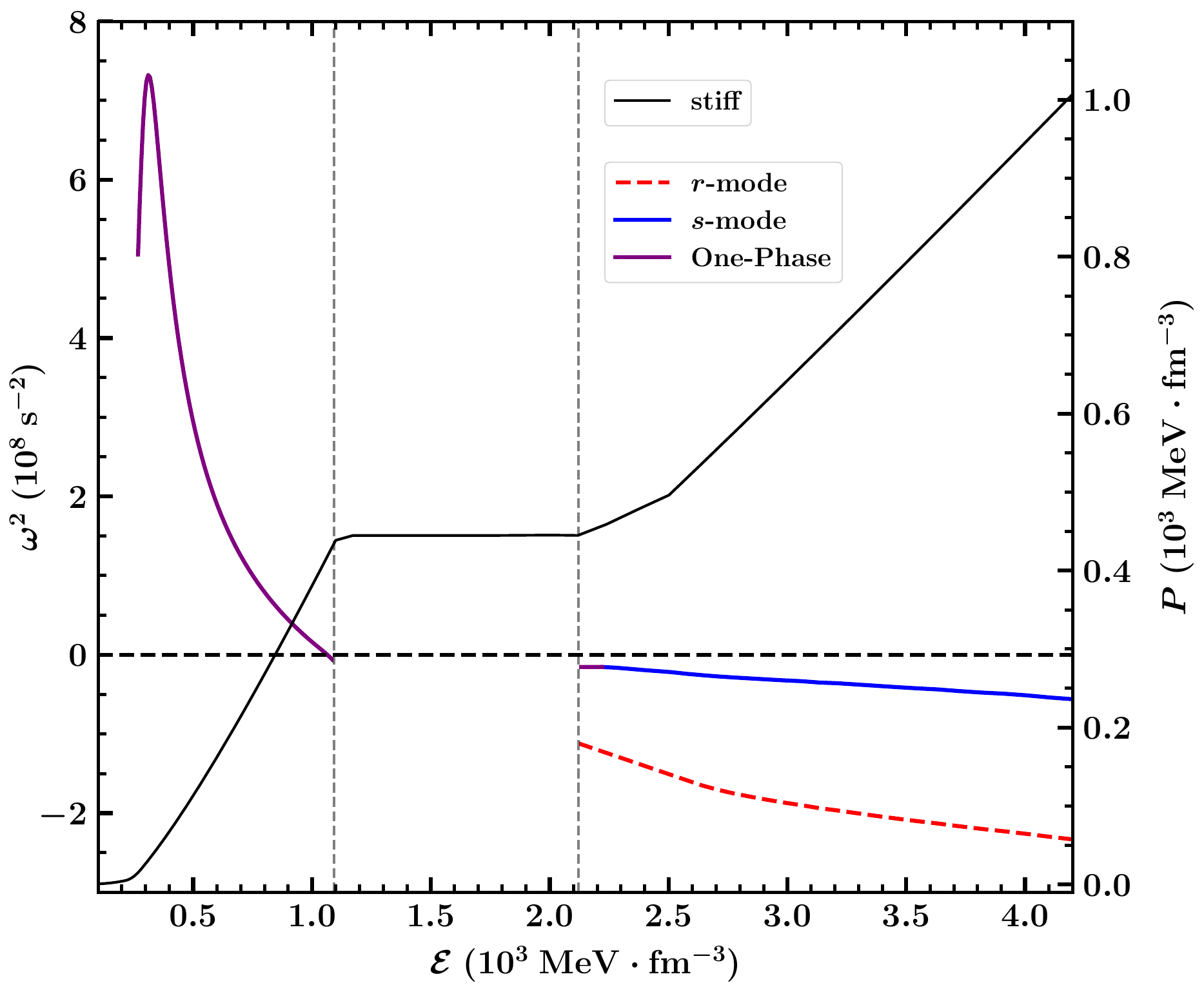}
    \includegraphics[width=0.5\linewidth]{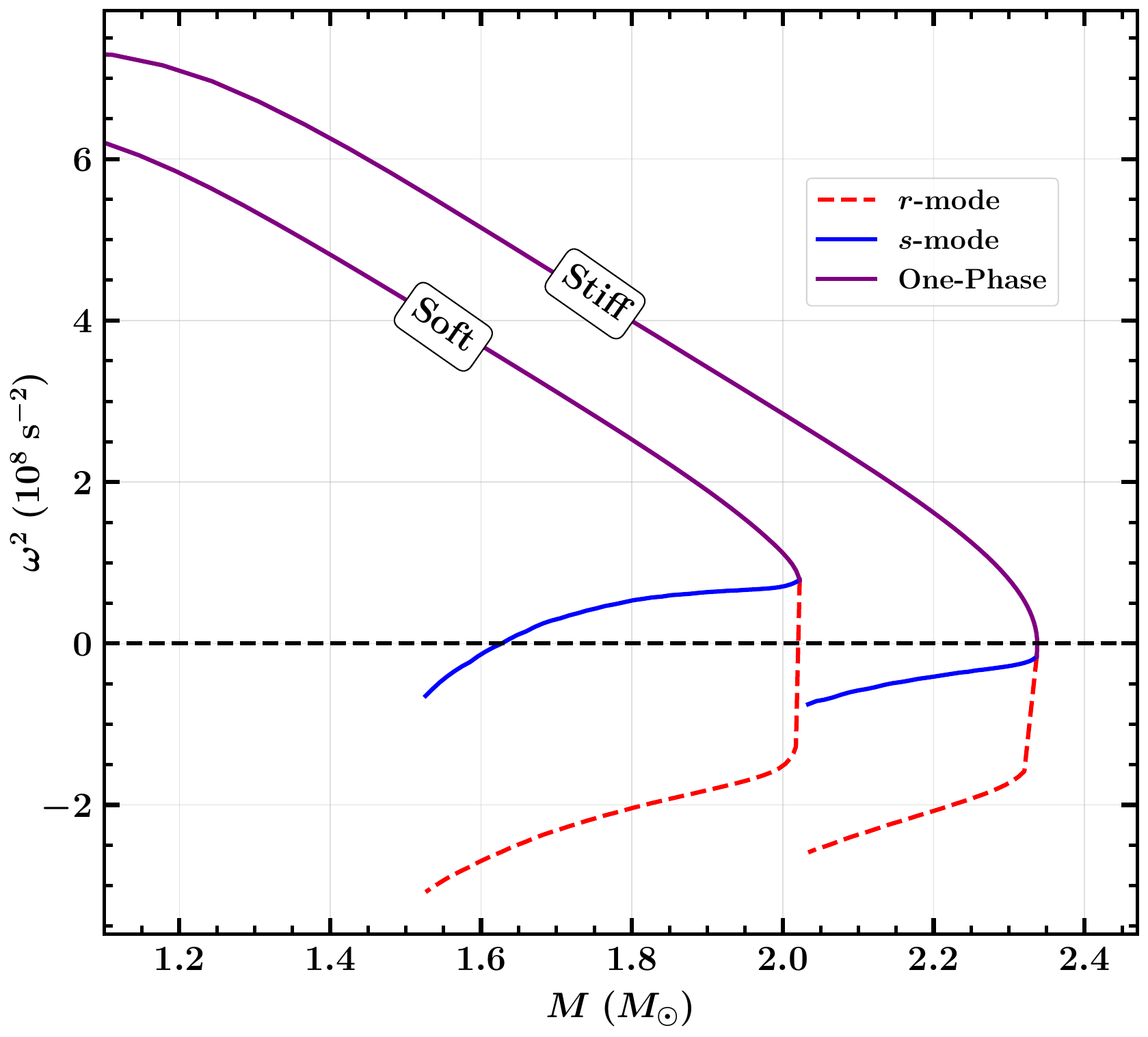}
    \caption{\textit{Top Left:} Squared $f$-mode frequency as a function of NS's central energy density corresponding to soft HQPT EOS for isotropic case. \textit{Top Right:} Squared $f$-mode frequency as a function of NS's central energy density corresponding to stiff HQPT EOS for isotropic case. The black line represents the relation between $\mathcal{E}$ and $P$ of the respective EOS. It is overplotted on twin axis, in order to highlight its density discontinuity. \textit{Bottom:} Squared $f$-mode frequency as a function of NS's mass corresponding to respective EOS for isotropic case. The purple line illustrates the squared $f$-mode frequency during the hadronic phase preceding the phase transition. The blue line denotes the $f$-mode associated with slow phase transitions, also referred to as $s$- or slow-modes. The red-dashed line signifies the $f$-mode linked to rapid phase transitions, also known as $r$- or rapid-modes.}
    \label{fig:rmodes}
\end{figure*}
\begin{table*}
    \centering
    \caption{NS parameters data obtained from standard criterion stability curve at stability limit ($\frac{\partial M}{\partial \rho_c } = 0 $) where maximum mass is attained. $\rho_c$ in ($\mathrm{10^{18} \ kg/m^3}$), $M$ in ($M_\odot$), $R$ in (km), $I$ in ($\mathrm{10^{45} \ g/cm^3}$), $k_2$ and $\Lambda$ are dimensionless, $f_o$ in ($\mathrm{kHz}$).}
    \renewcommand{\arraystretch}{1.2}
    \scalebox{1.1}{
    \begin{tabular}{cccccccccccccccc}
        \hline \hline & \multicolumn{7}{c}{$\mathrm{BSk21}$} & & \multicolumn{7}{c}{$\mathrm{SLY4}$} \\
        \cline { 2 - 8 } \cline { 10 - 16 } ${\beta_\mathrm{BL}}$ & $\rho_c$ & $M$ & ${R}$ & $I$ & ${k_2}$ & ${\Lambda}$ & ${f_o}$ && $\rho_c$ & $M$ & ${R}$ & $I$ & ${k_2}$ & ${\Lambda}$ & ${f_o}$ \\
        \hline 
        0 & 2.271 & 2.278 & 11.056 & 2.643 & 0.018 & 4.717 & {0.0} && 2.853 & 2.050  & 9.984 & 1.908 & 0.017 & 4.360 & {0.0} \\
        0.1 & 2.179 & 2.333 & 11.156 & 2.807 & 0.014 & 3.297 & 0.438i && 2.741 & 2.098 & 10.068 & 2.023 & 0.013 & 3.228 & 0.483i \\
        0.2 & 2.089 & 2.390 & 11.264 & 2.987 & 0.012 & 2.554 & 0.602i && 2.630 & 2.149 & 10.156 & 2.149 & 0.011 & 2.519 & 0.664i \\
        0.3 & 2.001 & 2.451 & 11.376 & 3.184 & 0.010 & 2.054 & 0.718i && 2.522 & 2.202 & 10.248 & 2.288 & 0.010 & 2.020 & 0.792i \\
        0.4 & 1.915 & 2.515 & 11.496 & 3.403 & 0.009 & 1.682 & 0.807i && 2.416 & 2.258 & 10.348 & 2.441 & 0.009 & 1.650 & 0.891i \\
        0.5 & 1.831 & 2.583 & 11.620 & 3.645 & 0.008 & 1.392 & 0.878i && 2.312 & 2.317 & 10.456 & 2.609 & 0.008 & 1.363 & 0.970i \\
        \hline  \hline \\
    \end{tabular}}
    \scalebox{1.1}{
    \begin{tabular}{cccccccccccccccc}
        \hline \hline & \multicolumn{7}{c}{$\mathrm{Soft}$} & & \multicolumn{7}{c}{$\mathrm{Stiff}$} \\
        \cline { 2 - 8 } \cline { 10 - 16 } ${\beta_\mathrm{BL}}$ & $\rho_c$ & $M$ & ${R}$ & $I$ & ${k_2}$ & ${\Lambda}$ & ${f_o}$ && $\rho_c$ & $M$ & ${R}$ & $I$ & ${k_2}$ & ${\Lambda}$ & ${f_o}$ \\
        \hline 
        0 & 2.574 & 2.022 & 11.904 & 2.437 & 0.041 & 27.592 & 1.551 && 1.887 & 2.337 & 11.884 & 3.073 & 0.025 & 8.141 & 0.0 \\
        0.1 & 2.498 & 2.089 & 11.972 & 2.596 & 0.029 & 17.243 & 1.346 && 1.824 & 2.400 & 12.005 & 3.281 & 0.018 &  5.463 & 0.520i \\
        0.2 & 2.421 & 2.159 & 12.036 & 2.769 & 0.023 & 11.911 & 1.117 && 1.744 & 2.467 & 12.128 & 3.509 & 0.015 &  4.084 & 0.615i \\
        0.3 & 2.359 & 2.231 & 12.100 & 2.957 & 0.019 &  8.603 & 0.845 && 1.667 & 2.538 & 12.259 & 3.760 & 0.013 &  3.202 & 0.709i \\
        0.4 & 2.310 & 2.307 & 12.160 & 3.160 & 0.016 &  6.357 & 0.465 && 1.592 & 2.613 & 12.394 & 4.041 & 0.011 &  2.573 & 0.795i \\
        0.5 & 2.261 & 2.387 & 12.220 & 3.383 & 0.014 &  4.744 & 0.493i && 1.518 & 2.693 & 12.539 & 4.357 & 0.010 &  2.096 & 0.861i \\
        \hline  \hline \\
    \end{tabular}}
    \label{tab:standard_criterion_table}
\end{table*}
\begin{figure*}
    \centering
    \includegraphics[width=0.49\linewidth]{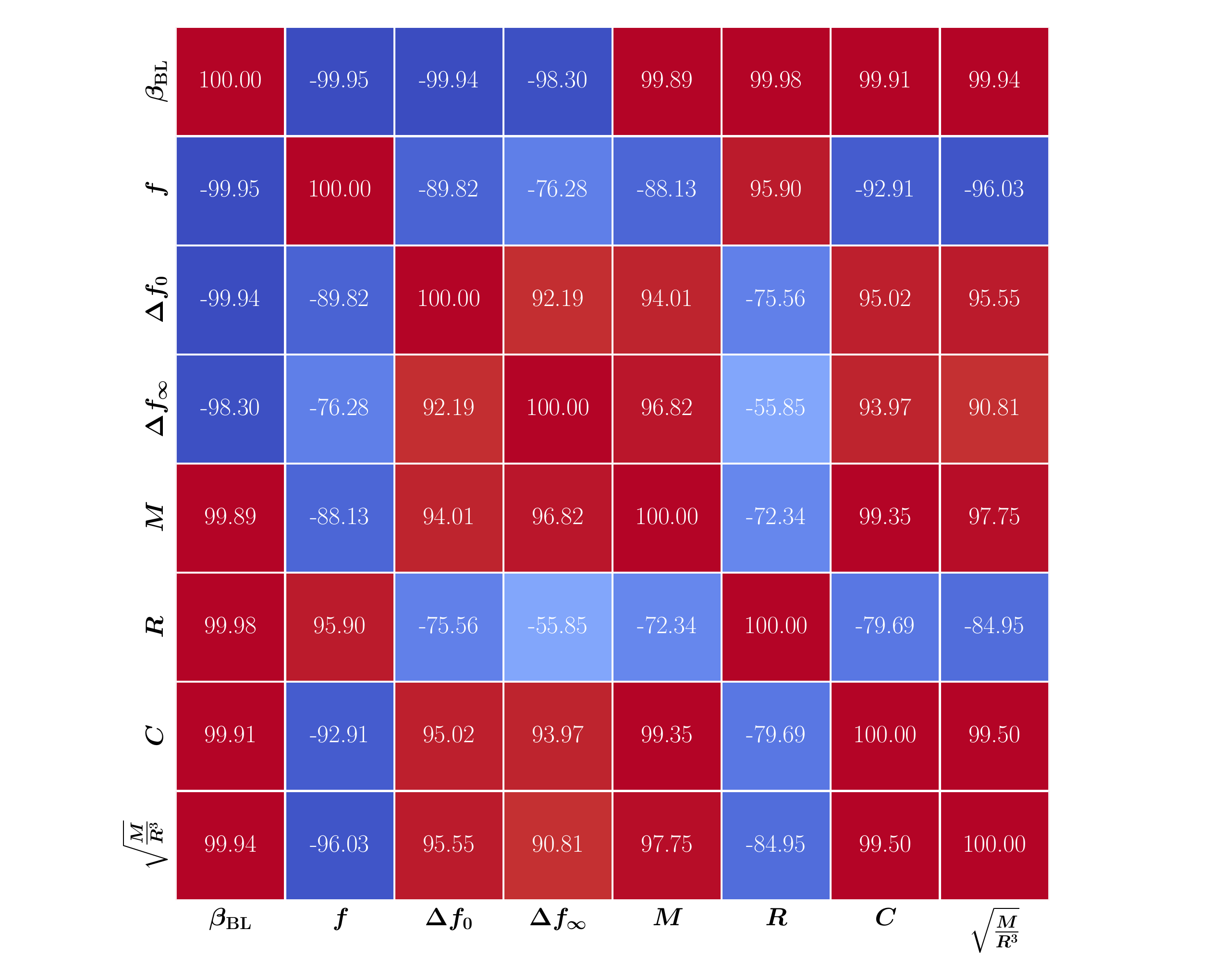}%
    \includegraphics[width=0.49\linewidth]{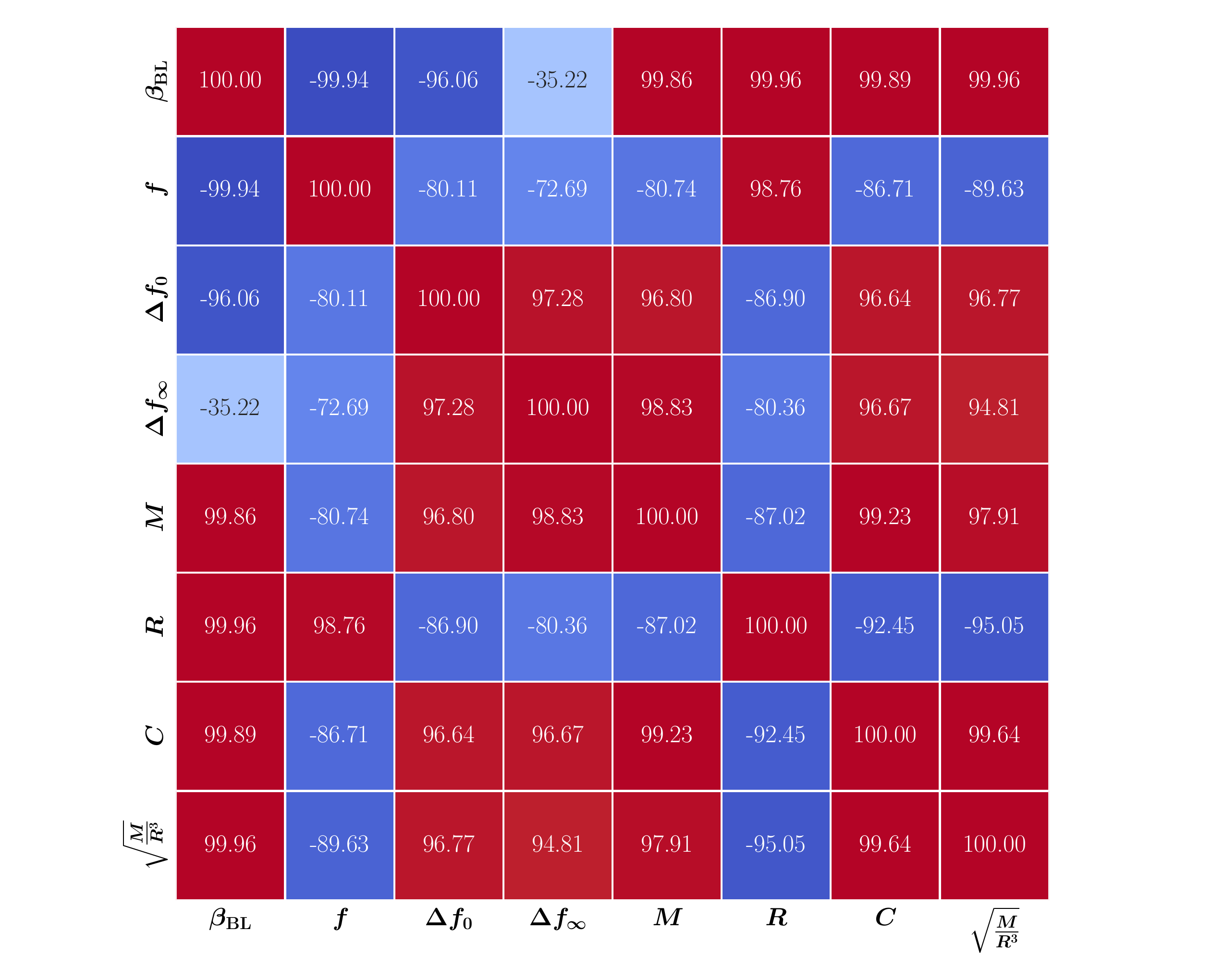}
    \includegraphics[width=0.49\linewidth]{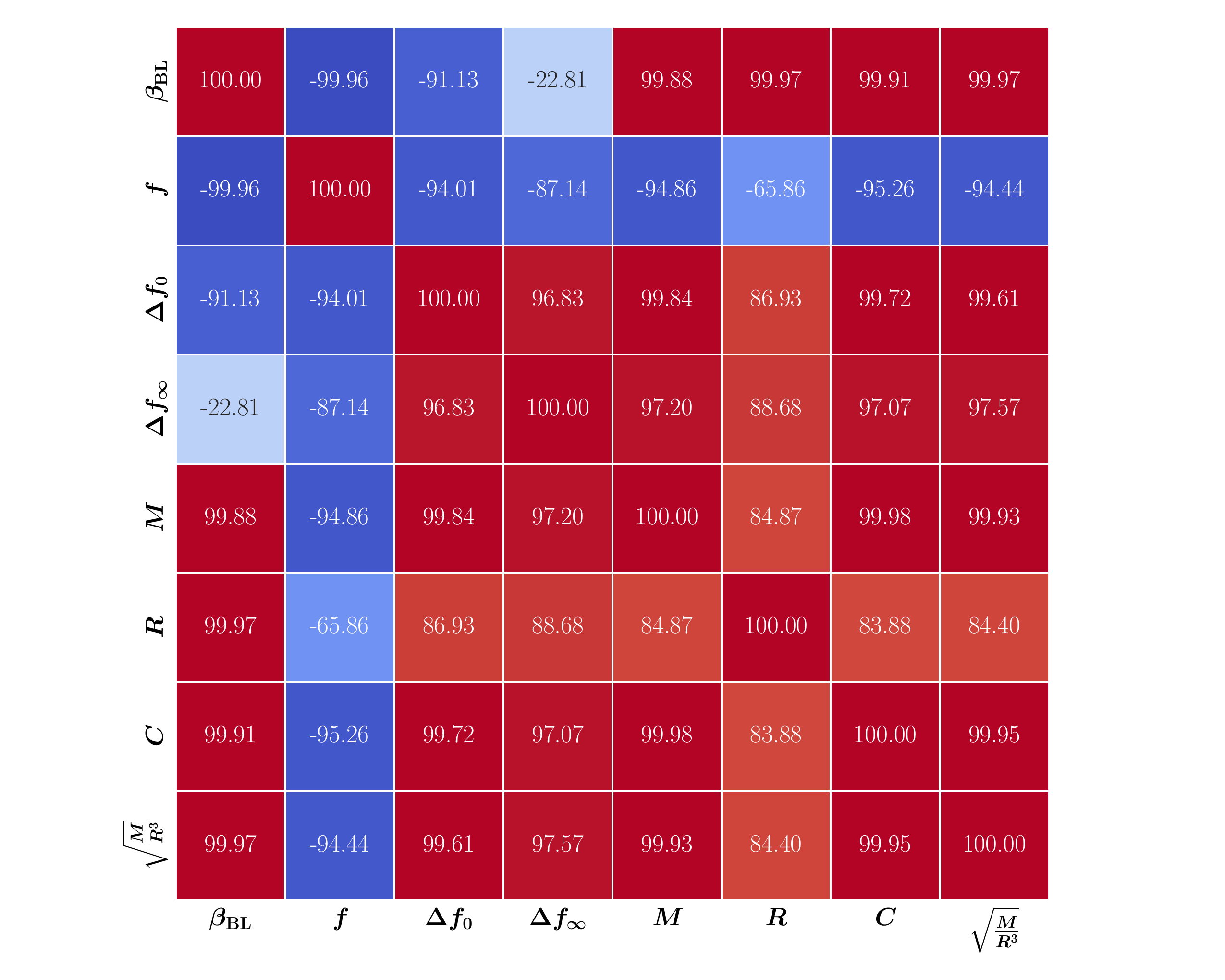}%
    \includegraphics[width=0.49\linewidth]{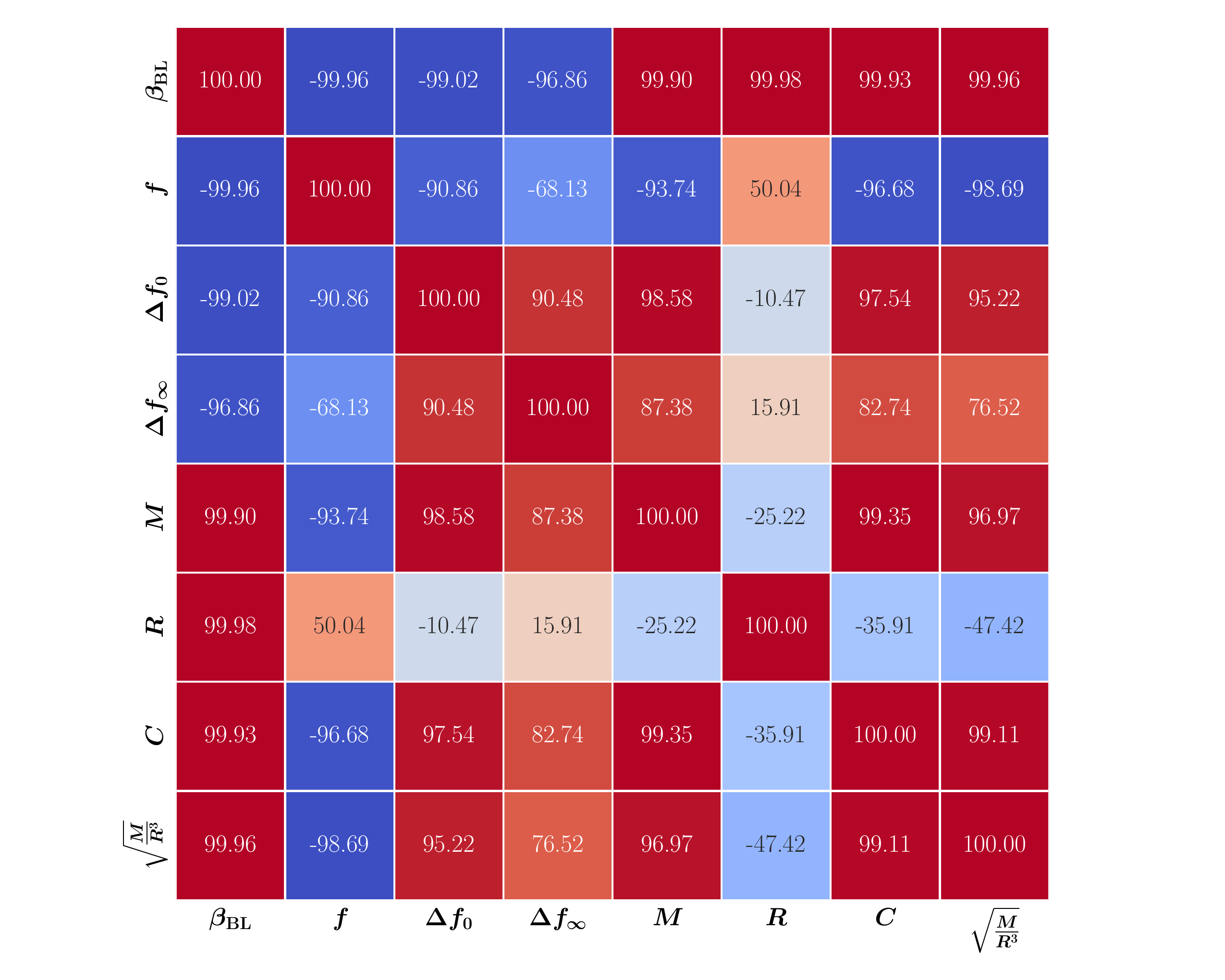}
    \caption{Correlation of NS parameters with anisotropy parameter for NSs ranging between $ (1 - 2) \ M_\odot $. \textit{Upper Left}: Corresponds to BSk21 EOS. \textit{Upper Right}: Corresponds to SLY4 EOS. \textit{Lower Left}: Corresponds to Soft HQPT EOS. \textit{Lower Right}: Corresponds to Stiff HQPT EOS}
    \label{fig:corr_plot}
\end{figure*}

The main objectives of this work are to examine the behavior of NS properties in an anisotropic background following BL-Model and to investigate the stability and non-adiabatic gravitational collapse of these anisotropic NSs. This study considers four EOSs representing hadronic and hadron-quark phase transitions for the NSs. The key findings of the present work are:

\begin{enumerate}
\item  {We studied the behavior of NS properties in an anisotropic background following the BL-Model. Along with pure hadronic EOSs, we also used phase transition EOSs in order to analyze how phase transition will affect NS properties. In the case of HQPT EOSs, we have meticulously considered the essential junction conditions to prevent the formation of {singularities resulting from energy discontinuities in our calculations.}}

\item We investigated the stability of anisotropic NSs using radial oscillation and found out that $f$-mode frequency corresponding to HQPT EOS tends to remain constant during phase transition and then drops once quark dominates the core, thus increasing NS's stability limit. For this NSs following HQPT EOS tends to remain stable even after crossing the maximum mass configuration. We also analyzed the effects of anisotropy on the large frequency separation for systems with and without crust, which is used in asteroseismology in order to study the interior composition of NSs. The frequency separation plot of hadronic EOSs without crust had a linear relation with frequency, but we could observe non-linear relation in the case of HQPT EOSs due to the abrupt changes of its adiabatic index at higher densities. We found out that the microphysics of the interior of NS (EOS, anisotropy) is imprinted on the large frequency separation. Detectability of radial oscillation along with analysis of consecutive frequency separation could shed some light on both micro as well as macro properties of NSs.

\item We examined the evolution of anisotropic NSs undergoing non-adiabatic gravitational collapse by filtering out unstable stars beyond the stability limit imposed by radial stability or $f$-mode curve. The collapse process was modeled based on ref. \cite{Pretel2020}. As a result of a radial heat flow, all key physical parameters abruptly change close to the event horizon's development. To better visualize this process, we performed a contour simulation, as shown in Fig. \ref{fig:ns-params_time_evolution}. We also examined the effects of phase transition in the evolution of key quantities of NS during the collapse. Black-hole time usually decreases with increasing central energy density or decreasing eigenfrequency for hadronic EOS, but it increases in the case of HQPT EOSs due to dominance quark in the core of such NSs. 

\item {Our research has led us to propose a viable method for detecting these immensely energetic gravitational collapse events. One effective approach involves utilizing optical photometry, allowing us to directly observe the abrupt surge in luminosity and energy emission. By utilizing observational data collected from a collapsing neutron star, one can determine the Full Width at Half Maximum (FWHM), denoted as $\Delta T$, of the luminosity-time plot. This FWHM measurement enables the calculation of crucial parameters characterizing the static neutron star prior to its collapse, including Compactness, Mass, Radius, etc.}
\end{enumerate}

Our understanding of the detectability of gravitational collapse can be significantly improved by observing the emission of GWs. In this study, we focused on the case of an unstable anisotropic NS undergoing radial collapse, where no GW is emitted during the process of gravitational collapse. However, considering more realistic scenarios before the onset of gravitational collapse, such as the presence of mountains in NSs \cite{Haskell}, rapidly rotating and magnetized NSs \cite{das2023detection}, and the deformation of NSs due to companion celestial objects, can heavily impact the emission of GW during the process of gravitational collapse. Moreover, during the binary NS inspiral-merger-ringdown phase \cite{Baiotti2022}, if one of the stars becomes unstable and undergoes gravitational collapse, it would emit GW that can be detected by current and future GW observatories like LIGO, Virgo, and LISA. Hence, studying these scenarios can not only improve our understanding of gravitational collapse but also lead to the discovery of new astrophysical phenomena.

\section{Acknowledgments}

B.K. acknowledges partial support from the Department of Science and Technology, Government of India with grant no. CRG/2021/000101. We express our sincere gratitude to Dr. Luciano Rezzola and Christian Ecker, who generously provided us with the phase-transition Equations of States.

\appendix

\section{Rapid versus Slow Modes}
\label{rs}
Energy density discontinuity in phase transition EOSs can have huge impact on radial modes of corresponding hybrid NSs. The rate of phase transition plays a crucial role is determining radial modes. In the presence of perturbations, the transition radius can either oscillate or remain static in the limiting cases of conversion speed: either slow, hence accompanying the fluid movement at the transition radius, or rapid, maintaining a static state. The radial modes of hybrid NSs obtained during a rapid phase transition are referred to as $r$- or rapid-modes, while for a slow phase transition, they are denoted as $s$- or slow-modes. Fig. \ref{fig:rmodes} represents the fundamental radial modes corresponding to soft and stiff variants of HQPT EOSs \cite{demircik2021dense} for isotropic NSs used in this study. The upper panel represents the relation between squared f-mode frequencies and NS's central energy density. The relation between $\mathcal{E}$ and $P$ for the corresponding EOS is overplotted with respect to twin axis in order to highlight the discontinuity in energy density. The Lower panel represents the relation between squared f-mode frequencies and NS's mass. The purple lines depicts the radial f-mode of NSs having central density below the density where phase transition begins. These NSs follow only the hadronic segment of the HQPT EOS, making them inherently one-phase or single-phase entities. The blue line represents the fundamental slow radial modes ($s$-mode) obtained by employing Eq. \eqref{eq:slow} while the red-dashed line depicts the fundamental rapid radial modes ($r$-mode) obtained by employing Eq. \eqref{eq:rapid}. The $s$-mode corresponding to soft HQPT EOS remains real (or the squared frequency is positive) till zero eigenfrequency is achieved, which suggests that the hybrid NSs tends to remain stable way beyond the maximum mass configuration. While the $r$-modes are imaginary as the corresponding squared frequency suffers a sharp drop just after phase transition and becomes negative, which suggests that all the hybrid NSs corresponding to soft HQPT EOS are unstable. In the case of stiff HQPT EOS, the squared frequencies of $r$-modes are lower than the $s$-mode but are negative (frequency is imaginary), suggesting that all hybrid stars are unstable for both slow and rapid phase transition cases. This happens because the zero eigenfrequency is attained during one-phase state before the phase transition occurs. Hence when we consider rapid phase transition, no hybrid NSs are found to be stable for either of the HQPT EOS, but while considering slow phase transition, hybrid NSs tend to remain stable till zero eigenfrequency is attained for soft variant of HQPT EOS.
\section{Stability Conditions}

As stated before, it has already been shown in recent literature such as ref. \cite{singh2022study,Takami,Weih,Pereira_2018,Rau_2023,Pretel2020} that the standard criterion for stability ($\frac{\partial M}{\partial \rho_c } < 0$) cannot be used in conjunction with frequency calculations to predict the onset of instability in anisotropic NS systems. This is because there is no correspondence between the stability limit obtained from squared frequencies of the fundamental mode and the star configurations that produce the maximum mass. For generality, we have also explored both $M(\rho_c)$ and $f$-mode method in order to gain some insights on the proper stability bounds imposed by both methods. Table \ref{tab:standard_criterion_table} lists the NS parameters for which the stability limit is reached, indicated by $\frac{\partial M}{\partial \rho_c } = 0$, for different $\beta_\mathrm{BL}$ corresponding to respective EOSs. These values represent the maximum mass configurations for different levels of $\beta_\mathrm{BL}$. In the $M(\rho_c)$ method, the first maximum on the $M(\rho_c)$ curve corresponds to a critical central density $\rho_c = \rho_{critical}$. This density marks the boundary between a stable family of stars and those that are unstable to gravitational collapse. On the $M(\rho_c)$ curve, the unstable branch of stars lies beyond the critical density where $dM/d \rho_c < 0$. 

For the hadronic EOSs (BSk21 \& SLY4), the $f$-mode frequency tends to zero for the isotropic case, and it becomes imaginary (or $\omega^2$ becomes negative) for positive anisotropic cases. This implies that the stability limit imposed by radial oscillations for anisotropic NS is reached before the maximum mass configuration, meaning that the critical central density obtained from the standard criterion $M(\rho_c)$ method does not always indicate the onset of instability. Therefore, it is clear that anisotropy has an impact on stellar stability, and additional conditions beyond the standard criterion are needed to determine the limits of stability. The same can be observed for Stiff HQPT EOS. But for Soft HQPT EOS, the $f$-mode frequency for the isotropic case corresponding to the maximum mass configuration is about 1.5 kHz. This suggests that maximum mass is attained quite earlier than reaching the zero eigenfrequency mode and that the star remains stable for a long range of central energy density.

\section{NS Parameters \& Anisotropy Correlation}
\label{comax}

To support our statements about effects of anisotropy on consecutive frequency difference and how along with other NS parameters anisotropy is also imprinted on large separation, we have calculated the correlation matrix among various neutron star properties such as f-mode frequency ($f$), large frequency difference ($\Delta f_0$), Mass ($M$), Radius ($R$), Compactness ($C$) \& root mean density ($\sqrt{M/R^3}$), along with anisotropy parameter ($\beta_{\mathrm{BL}}$). In Fig. \ref{fig:corr_plot}, we present the correlation matrices for distinct NS properties specific to the respective equations of state (EOSs). These correlations were meticulously calculated for a series of neutron stars with masses ranging from 1.0 to 2.0 solar masses ($M_\odot$) and anisotropy parameters spanning the interval $0 < \beta_{\mathrm{BL}} < 0.5$.

\bibliographystyle{apsrev4-1}
\bibliography{GC}
\end{document}